\documentclass[aps,prd,reprint,superscriptaddress,nofootinbib,floatfix,longbibliography]{revtex4-1}
\usepackage[dvipsnames]{xcolor}
\usepackage{amssymb,amsmath,amsfonts,bm,slashed,soul,ragged2e,graphicx,epstopdf,hyperref,array}
\usepackage[utf8]{inputenc}
\usepackage{caption,subcaption}
\enlargethispage{\baselineskip}
\hypersetup{colorlinks,linkcolor={blue},citecolor={blue},urlcolor={blue}}
\DeclareCaptionJustification{justified}{\justifying}
\everymath{\displaystyle} 
\usepackage{mathrsfs}

\newcommand{\re}{\text{Re}}
\newcommand{\nn}{{\nonumber}}

\newcommand{\beq}{\begin{equation}}
\newcommand{\eeq}{\end{equation}}
\newcommand{\bea}{\begin{eqnarray}}
\newcommand{\eea}{\end{eqnarray}}

\newcommand{\gsim}{\lower.7ex\hbox{$\;\stackrel{\textstyle>}{\sim}\;$}}
\newcommand{\lsim}{\lower.7ex\hbox{$\;\stackrel{\textstyle<}{\sim}\;$}}

\newcommand{\be}{\begin{equation}}
\newcommand{\ee}{\end{equation}}
\newcommand{\ba}{\begin{eqnarray}}
\newcommand{\ea}{\end{eqnarray}}

\newcommand{\cor}[1]{\ensuremath{\langle #1 \rangle}}
\newcommand{\vvir}{v_{\textrm{vir}}}

\newcommand{\vir}{\textrm{vir}}
\newcommand{\diag}{\textrm{diag}}
\newcommand{\EDM}{\textrm{EDM}}
\newcommand{\MDM}{\textrm{MDM}}
\renewcommand{\parallel}{\mathbin{\!/\mkern-5mu/\!}}
\newcommand{\T}{\textrm{T}}
\newcommand{\Ana}{\textrm{Ana}}
\newcommand{\mi}{\mathrm{i}}
\newcommand{\me}{\mathrm{e}}
\newcommand{\co}{\mathcal{O}}

\begin{document} 
	\title{Dissecting axion and dark photon with a network of vector sensors}

			\author{Yifan Chen$^{a}$}
			\author{Min Jiang$^{b,c,d}$}
			\author{Jing Shu$^{a,e,f,g}$}
			\author{Xiao Xue$^{h}$}
			\author{Yanjie Zeng$^{a,e}$}

\affiliation{
$^{a}$CAS Key Laboratory of Theoretical Physics, Institute of Theoretical
Physics,\\ Chinese Academy of Sciences, Beijing 100190, P.R.China\\
$^b$Hefei National Laboratory for Physical Sciences at the Microscale and Department of Modern Physics,
University of Science and Technology of China, Hefei, Anhui 230026, China\\
$^c$CAS Key Laboratory of Microscale Magnetic Resonance,
University of\\ Science and Technology of China, Hefei, Anhui 230026, China\\
$^d$Synergetic Innovation Center of Quantum Information and Quantum Physics,\\
University of Science and Technology of China, Hefei, Anhui 230026, China\\
$^e$School of Physical Sciences, University of Chinese Academy of Sciences, Beijing 100049, China\\
$^f$School of Fundamental Physics and Mathematical Sciences, Hangzhou Institute for Advanced\\ Study, University of Chinese Academy of Sciences, Hangzhou 310024, China\\
$^g$International Center for Theoretical Physics Asia-Pacific, Beijing/Hangzhou, China\\
$^h$II. Institute of Theoretical Physics, Universit\"at Hamburg, 22761 Hamburg, Germany
}
		
		\begin{abstract}
		We develop formalisms for a network of vector sensors, sensitive to certain spatial components of the signals, to identify the properties of a light axion or a dark photon background. These bosonic fields contribute to vector-like signals in the detectors, including effective magnetic fields triggering the spin precession, effective electric currents in a shielded room, and forces on the matter. The interplay between a pair of vector sensors and a baseline that separates them can potentially uncover rich information of the bosons, including angular distribution, polarization modes, source localization, and macroscopic circular polarization. Using such a network, one can identify the microscopic nature of a potential signal, such as distinguishing between the axion-fermion coupling and the dipole couplings with the dark photon.
\end{abstract}

\date{\today}

\maketitle

\section{Introduction}
Bosons in the sub-eV mass range can be natural dark matter candidates.
These non-relativistic bosonic fields behave like coherent waves within the correlation time and distance. Among these, axion dark matter \cite{Preskill:1982cy, Abbott:1982af, Dine:1982ah} is strongly motivated. As a pseudo-scalar, axion can solve the strong charge-conjugation and parity (CP) problem in quantum chromodynamics (QCD) \cite{Peccei:1977hh}. In addition to the QCD-axion, axion-like particles (ALPs) also appear generically  in theories with extra dimensions {\cite{Arvanitaki:2009fg}}. Dark photon dark matter \cite{Nelson:2011sf} is another popular candidate for the bosonic wave-like dark matter. Similar to the axion, it can originate from the string compactification \cite{Abel:2008ai,Goodsell:2009xc} and is produced from misalignment.

In addition to being non-relativistic dark matter, light bosons can also exist in other forms in the universe. For example, non-virialized cold streams \cite{OHare:2017yze,Foster:2017hbq,Knirck:2018knd}, dipole radiations from binary systems charged under a hidden $U(1)$ \cite{Krause:1994ar,Dror:2021wrl,Hou:2021suj}, and emissions from gravitational atoms  \cite{Baryakhtar:2020gao,East:2022ppo} contribute to highly anisotropic bosonic waves from a specific direction. Some 
relativistic degrees of freedom can remain in the current universe as well, forming a cosmological background \cite{Baumann:2016wac,Dror:2021nyr}.

The detection of these bosonic background fields is attracting a growing number of different experiments. A natural step forward is to construct a network of detectors.
Besides cross-checking and distinguishing the signals from spurious noise,
a multi-mode resonant system with $\mathcal{P T}$ symmetry can significantly enhance the signal power and the scan rate \cite{Li:2020cwh,Chen:2021bgy}. A pair of detectors with a long baseline can read the directional information of the dark matter \cite{Derevianko:2016vpm, Foster:2020fln} and opens a new window in multi-messenger astronomy \cite{Dailey:2020sxa}. These works are based on scalar-like observables, such as the time-derivative of the axion field in electromagnetic haloscopes \cite{Sikivie:1983ip, Sikivie:1985yu, Sikivie:2013laa, Chaudhuri:2014dla, Berlin:2019ahk,ADMX:2021nhd,CAPP:2020utb,HAYSTAC:2020kwv,McAllister:2017lkb,Salemi:2021gck}.

On the other hand, there are types of experiments sensitive to vector-like signals, including effective magnetic fields leading to spin precession \cite{Graham:2013gfa,Budker:2013hfa,Stadnik:2013raa,Barbieri:2016vwg,Abel:2017rtm,Stadnik:2017mid,jiang2021floquet,wu2019search,garcon2019constraints,Smorra:2019qfx,Mitridate:2020kly,Chigusa:2020gfs,aybas2021search,Jiang:2021dby,Kim:2021eye,Poddar:2021ose}, that can capture axion-fermion coupling or dipole couplings with a dark photon \cite{Graham:2017ivz}, effective electric currents from kinetic-mixing dark photons 
\cite{Chaudhuri:2014dla,Fedderke:2021aqo,Fedderke:2021rrm}, and forces brought from $U(1)_{B-L} / U(1)_{B}$ dark photons \cite{Graham:2015ifn,Pierce:2018xmy,Carney:2019cio,Guo:2019qgs}. Taking the axion-fermion coupling as an example,  in the non-relativistic limit of the fermion, one has vector-like coupling between the spatial derivative of the axion and the spin operator of the fermion,
\be g_{a} \partial_{\mu} a \bar{\psi} \gamma^{\mu} \gamma^{5} \psi \rightarrow g_{a} \vec{\nabla} a \cdot \vec{\sigma},\ee
where $a$ and $\psi$ are the axion and the fermion field, respectively, with coupling constant $g_{a}$, and $\vec{\sigma}$ is the spin operator of the fermion. The vector-like signal in this case is the axion wind/gradient $\vec{\nabla} a$, which can be detected through nuclear magnetic resonance \cite{Graham:2013gfa,Budker:2013hfa,garcon2019constraints,aybas2021search}, Floquet masers \cite{jiang2021floquet}, comagnetometers \cite{Abel:2017rtm,wu2019search,bloch2020axion}, magnons \cite{Barbieri:2016vwg,Mitridate:2020kly,Chigusa:2020gfs}, a Penning trap \cite{Smorra:2019qfx}, or spin-based amplifiers \cite{Jiang:2021dby,su2021search}.

An outstanding question to ask then is, what information can one extract by correlating a pair of vector sensors? Since a single vector sensor can in principle probe any spatial direction by manipulating the configuration of the detector, one expects the interplay between the two sensitive directions and the baseline that separates them to bring a correlation matrix
\begin{equation}
    \cor{(\vec{\mathcal{O}}(t_1,\vec{x}_1) \cdot \hat{l}_1)(\vec{\mathcal{O}}(t_2,\vec{x}_2) \cdot \hat{l}_2)},
    \label{veccorr}
\end{equation}
where $\hat{l}_1, \hat{l}_2$ are the sensitive directions of the two vector sensors and $\vec{\mathcal{O}}$ are the vector-like signals recorded at different times and locations by the corresponding detectors. $\cor{\cdots}$ denotes the ensemble average on the bosonic fields. As we will discuss, the correlation function (\ref{veccorr}) can bring important information about the bosonic background, including angular distribution, polarization modes, coupling types, the exact direction of the origin, and macroscopic circular polarization. With these, one can easily identify the microscopic nature of a potential signal.
Notice that there are already many vector sensors operating at the same time in The Global Network of Optical Magnetometers for Exotic physics (GNOME)  \cite{Pospelov:2012mt, Pustelny:2013rza, Afach:2018eze, Afach:2021pfd}.

The layout of the paper is as follows. In Sec.\,\ref{VS}, we review several types of coupling that contribute to the vector-like signals in terms of stochastic wave functions of the bosons, and we introduce the linear and circular polarization basis of the vector sensors.
In Sec.\,\ref{S3}, we generalize the scalar sensor interferometry to a network of vector sensors where the two-point correlations of the signals are equipped with two additional labels of spatial component, and we discuss the universal dipole angular correlation for isotropic sources.
In Sec.\,\ref{S4}, we discuss the identifications of longitudinal and transverse modes for incoming and isotropic sources, and different types of dark matter contributing to vector-like signals.
In Sec.\,\ref{SL}, we calculate the angular resolution for incoming sources with different polarization modes when the vector sensors are set to the optimal directions in terms of the baseline and the incoming source.
In Sec.\,\ref{CP}, we show how to identify a macroscopic circular polarization for an isotropic source.
Section \ref{conclusion} contains the conclusion and  future prospects.

\section{Vector Sensor Response to Axion Gradient and Dark Photon}\label{VS}

\subsection{Non-relativistic Dipole Coupling}\label{NRDC}

We first consider the dipole couplings between light bosons to the spins of fermions.
In the non-relativistic limit of the fermions, these dipole couplings can be expressed universally as an effective vector-like coupling with the fermion's spin operator $\vec{\sigma}$,
\begin{equation}
    H= \alpha \, \vec{\mathcal{O}}\cdot \vec{\sigma}.
    \label{hamD}
\end{equation}
where $\alpha$ is the coupling constant and $\vec{\mathcal{O}}$ is the vector-like signals probed.

Up to dimension-five operators are considered, which become the form of Eq.\,(\ref{hamD}) in the non-relativistic limit of the fermions, including axion-fermion coupling,
\begin{align}
    g_a \partial_{\mu} a \bar{\psi} \gamma^{\mu} \gamma^{5} \psi &\rightarrow \vec{\mathcal{O}}_a=\vec{\nabla} a, \label{axiongradient}\end{align}
    and dipole interactions between dark photon and the fermions
    \begin{align}
    g_A V_{\mu} \bar{\psi}\gamma^{\mu} \gamma^5 \psi &\rightarrow \vec{\mathcal{O}}_A=\vec{V}, \label{axialvec}\\
    g_{\MDM} V_{\mu \nu} \bar{\psi} \sigma^{\mu \nu} \psi &\rightarrow \vec{\mathcal{O}}_{\MDM}=\vec{\nabla}\times \vec{V},\label{MDMLH}\\
      g_{\EDM} V_{\mu \nu} \bar{\psi}\sigma^{\mu \nu} i \gamma^5 \psi &\rightarrow \vec{\mathcal{O}}_{\EDM}=\partial_0 \vec{V} - \vec{\nabla} V^0,\label{EDMLH}
\end{align}
where $g_a$, $g_A$, $g_{\EDM}$, and $g_{\MDM}$ are the coupling constants for the axion, axial-vector, electric dipole momentum (EDM) and magnetic dipole momentum (MDM) dark photon respectively. $a$ is the axion field, and $V_\mu$ and $V_{\mu\nu}$ are the vector fields and their field strength, respectively. $\sigma^{\mu \nu}$ is the anti-symmetric tensor constructed from Dirac matrix $\gamma^\mu$.
Notice that in the case of the axion gradient coupling (\ref{axiongradient}), $\vec{\co}$ behaves as a longitudinal mode in terms of the spatial momentum. In contrast, only transverse modes of the MDM dark photon contribute to the interaction in Eq.\,(\ref{MDMLH}). 
For the axial vector, all the polarization modes take part in the interaction in Eq.\,(\ref{axialvec}).
Interactions with the EDM dark photon differ between the relativistic and the non-relativistic limits of the dark photon in Eq.\,(\ref{EDMLH}). In the former case, only the transverse modes interact, while all contribute to the interaction when being non-relativistic. 

Operators with dimension higher than 5 always become one of the forms above in the non-relativistic limit. For example, an anapole dark photon, with interaction $g_{\Ana} (\partial^{\nu}V_{\mu \nu}) \bar{\psi}\gamma^{\mu} \gamma^5 \psi$, takes the same form as the axial-vector in Eq.\,(\ref{axialvec}) when the dark photon is on-shell.

For more general cases of the dark photon, one can also use Eq.\,(\ref{hamD}) to parametrize the detector response for a certain spatial direction. For example,  the $U(1)_{B-L} / U(1)_{B}$ dark photon leads to an additional force that accelerometers, optomechanical systems, or astrometry can sense \cite{Graham:2015ifn,Pierce:2018xmy,Carney:2019cio,Guo:2019qgs}. The kinetic-mixing dark photon induces an effective current which can be captured by LC circuits in a shielded room \cite{Chaudhuri:2014dla} or read from the geomagnetic field \cite{Fedderke:2021aqo,Fedderke:2021rrm}. In the former case, the force signal shares the same form as the one of the EDM in Eq.\,(\ref{EDMLH}), while the effective current is proportional to the spatial component of the dark photon wave functions.

\subsection{Vector-like Signals from Axion and Dark Photon Background}

Bosonic fields can exist in the universe with different kinds of momentum distribution. For example, axion or dark photon with mass $m < \mathcal{O} (1)$ eV can be cold dark matter candidates \cite{Preskill:1982cy, Abbott:1982af, Dine:1982ah, Nelson:2011sf}, behaving like coherent waves within the correlation time and distance that are both determined by virial velocity. A cosmological axion \cite{Baumann:2016wac, Dror:2021nyr} and a dark photon can also exist with the spectrum dependent on the production mechanism and the universe's thermal history. Furthermore, one expects bosonic waves to come from a specific direction towards the Earth,  such as non-virialized substructure of the dark matter \cite{OHare:2017yze, Foster:2017hbq, Knirck:2018knd}, dipole radiations from binary systems charged by hidden $U(1)$ \cite{Krause:1994ar,Dror:2021wrl,Hou:2021suj}, and emissions from strongly self-interacting gravitational atoms \cite{Baryakhtar:2020gao,East:2022ppo}. In this subsection, we construct the bosonic backgrounds and the corresponding vector-like signals using effective polarization modes.

Following \cite{Foster:2017hbq,Derevianko:2016vpm,Foster:2020fln,Guo:2019ker} and the relativistic generalization in \cite{Dror:2021nyr}, one expands general scalar or vector fields with $N_b$ non-interacting waves
\ba
	a(t,\vec{x}\,) &=& \sum_{j=1}^{N_b}\sqrt{\frac{2\rho}{N_b \bar{\omega}\omega_j}} \cos \left[ \omega_j t-\vec{p}_j \cdot \vec{x}+\alpha_j \right],\label{Sa}\\
    \vec{V}(t,\vec{x}\,) &=& \sum_{j=1}^{N_b}\sqrt{\frac{2\rho}{N_b \bar{\omega}\omega_j}}\re \left[ \me^{- \mi \left( \omega_j t-\vec{p}_j \cdot \vec{x}+\alpha_j \right)} \ \vec{\epsilon}_{\beta}^{\,j} \right],
    \label{Vector Field}
\ea
where momentum $\vec{p}_j$ are  random variables drawn from the momentum distribution  $f(\vec{p}\,)$, $\omega_j=\sqrt{m^2+ p_j^2}$ are the energy in which $p_j \equiv |\vec{p}_j|$, $\bar{\omega} \equiv \int d^3  \vec{p} \, f(\vec{p}) \, \omega$ is the average energy, and $\rho$ is the local energy density.
$\alpha_j$ are random phases uniformly distributed within $[0,2\pi)$ for a stochastic background that we focus on in this study. $\beta$ are discrete random variables following the probabilities of different polarization modes $\vec{\epsilon}_\beta$. 

Generally, a massive vector field contains three degrees of freedom.  In the unitary gauge, these include the longitudinal mode
\begin{equation}
    \vec{\epsilon}_{0} = \frac{\omega}{m} \hat{e}_{p} \equiv \frac{\omega}{m} \left(\sin \theta \cos \phi,\sin \theta \sin \phi,\cos \theta\right)^{\T},
\end{equation}
and two transverse ones in the linear polarization basis,
\begin{align}
    \vec{\epsilon}_{H}  = \hat{e}_{\theta} &\equiv \left(\cos \theta \cos \phi,\cos \theta \sin \phi,-\sin \theta \right)^{\T}, \\
     \vec{\epsilon}_{V} = \hat{e}_{\phi}  &\equiv \left(-\sin \phi,\cos \phi,0 \right)^{\T},
\end{align}
where $\hat{e}_{p}, \hat{e}_{\theta}$ and $\hat{e}_{\phi}$ are three orthogonal unit directional vectors. Alternatively, one can express the transverse ones using the circular polarization basis
\be
    \vec{\epsilon}_{R/L} = \frac{1}{\sqrt{2}} \left( \vec{\epsilon}_{H} \pm \mi\ \vec{\epsilon}_{V} \right).  \label{leftbasis}
\ee

Now the vector observables $\vec{\co}$ in Eq.\,(\ref{axiongradient}-\ref{EDMLH})  can be written as a sum of different polarization modes depending on a specific type of coupling, 
\ba
    \vec{\mathcal{O}}_a&=&\sum_{j=1}^{N_b}\, \frac{m\, p_j}{\omega_j}\, \Phi_{j}   \cos \left[ \Psi_j \right]  \, \vec{\epsilon}_{0}^{\,j} , \label{OAG}\\
   \vec{\mathcal{O}}_A&=&\sum_{j=1}^{N_b}\, \Phi_{j}  \,\re \left[ \me^{ -\mi \Psi_j} \, \vec{\epsilon}_{\beta}^{\,j} \right],\label{OAV}\\
 \vec{\mathcal{O}}_{\MDM}&=& \sum_{j=1}^{N_b}\, p_j \, \Phi_{j}\,  \re \left[ \me^{ -\mi \Psi_j} \ \vec{\epsilon}_{R/L}^{\,j} \right],\label{OMDM}\\
     \vec{\mathcal{O}}_{\EDM}&=& \sum_{j=1}^{N_b}\,
    \begin{cases}
      m\, \Phi_{j}  \,\re \left[ \me^{ -\mi \Psi_j} \, \vec{\epsilon}_{\beta}^{\, j}\right], & m \gg |p_j|, \\
      \omega_j\, \Phi_{j} \,\re \left[ \me^{ -\mi \Psi_j} \, \vec{\epsilon}_{R/L}^{\,j} \right],  & m \ll |p_j|,
    \end{cases}\nn\\\label{OEDM}
\ea
where one uses the constraint $\partial_{\mu} V^{\mu} =0 $ of the unitary gauge to express $V^0$ in Eq.\,(\ref{EDMLH}). $\Phi_{j} \equiv \sqrt{2\rho/N_b \bar{\omega} \omega_j}$ and $\Psi_j \equiv \omega_j t-\vec{p}_j \cdot \vec{x}+\alpha_j$ are defined for simplicity. One can thus treat the axion gradient effectively as the longitudinal-mode-only case of the dark photon.

\subsection{Circular and Linear Polarization Basis of Vector Sensors}

In this subsection, we start with a review on detecting effective magnetic fields from the spin precession and explain why the detectors are in the circular polarization basis, in analogy with the concept in radio astronomy. Thus one can transform it to the linear polarization basis with two systems polarized at opposite directions operating simultaneously, probing a specific spatial component of the vector-like signals on the transverse plane.

The couplings between the dark sectors and the spins of nucleons in Eq.\,(\ref{hamD}) can be detected through nuclear magnetic resonance \cite{Graham:2013gfa,Budker:2013hfa,garcon2019constraints,aybas2021search}, Floquet masers \cite{jiang2021floquet}, or spin-based amplifiers \cite{Jiang:2021dby,su2021search}.
 In such cases, the nucleons are initially polarized along a specific direction, and the signals $\vec{\co}$ on the transverse directions can be captured. Without loss of generality, taking the initial polarized spin $\vec{\sigma}_0$  along $+\hat{z}$, one can describe the dynamics of the spin operators $\vec{\sigma}$ using the Bloch equation
\begin{equation}
    \frac{d \vec{\sigma}}{dt}= \vec{\sigma} \times \left( \gamma \vec{B}_0 - \alpha \vec{\co} \right) - \frac{\vec{\sigma} - \vec{\sigma}_0}{T_r},\label{Blocheq}
\end{equation}
where $\gamma$ is the gyromagnetic ratio, $\vec{B}_0 = B_0 \hat{z}$ is the background magnetic field along $+\hat{z}$, and $T_r$ is the relaxation time of the system. One way to see the response to the signal is to decompose the transverse directions in terms of $\sigma_{\perp} \equiv \left(\sigma_x + \mi \sigma_y\right)/\sqrt{2}$, which leads to
\begin{equation}
     \left(\frac{d}{dt} + \mi \gamma B_0 + \frac{1}{T_r} \right) \sigma_\perp = - \mi \alpha \sigma_0 \left(\frac{ \co_x + \mi \co_y}{\sqrt{2}}  \right),
\end{equation}
where $\gamma B_0$ sets the resonant frequency, and $1/T_r$ denotes the dissipation. $\left( \co_x + \mi \co_y \right)/\sqrt{2}$ are the sensitive signals, which are clearly in the circular polarization basis. For transverse modes propagating along the $+\hat{z}$-axis, only the left-hand circular polarization mode $\epsilon_L$ can excite the detector. For the longitudinal mode, $\epsilon_0$, only the projection to the transverse plane gets a response, and one cannot distinguish its component along $\hat{x}$ and $\hat{y}$.

Two detectors with opposite initially polarized directions and the same magnetic field strength $B_0$ can transform the circular polarization basis into the linear one. More explicitly, one sums up simultaneously the signals from the detectors polarized along $+\hat{z}$ and $-\hat{z}$ with relative phase shift $\pi/2$ and gets the response to a specific spatial direction on the transverse plane,
\ba
     \left( \frac{d}{dt} + \mi \gamma B_0 + \frac{1}{T} \right) \left(  \frac{\sigma_{\perp}^{\uparrow} + \mi \sigma_{\perp}^\downarrow }{\sqrt{2}} \right) &=& - \mi \alpha \sigma_0  \co_{x}, \label{CtLX} \\ 
     \left( \frac{d}{dt} + \mi \gamma B_0 + \frac{1}{T} \right) \left(  \frac{\sigma_{\perp}^{\uparrow} - \mi \sigma_{\perp}^\downarrow }{\sqrt{2}} \right) &=&  \alpha \sigma_0  \co_{y}, \label{CtLY}
\ea
where $\sigma_\perp^{\uparrow/\downarrow}$ are the transverse responses for detectors polarized along $+\hat{z}$ and $-\hat{z}$ respectively.

This transformation of the response basis also applies to detectors using polarized electrons, such as the axion-magnon conversion  \cite{Barbieri:2016vwg,Mitridate:2020kly,Chigusa:2020gfs}. 
 For the $U(1)_{B-L} / U(1)_{B}$ dark photon or the kinetic-mixing dark photon, as mentioned in Sec.\,\ref{NRDC}, the detector responses to them are in the linear polarization basis.

\section{From the Scalar Sensor Interferometry to Vector Sensor Network}\label{S3}
This section starts with a review on the dark matter interferometry developed in \cite{Foster:2020fln}, where directional information can be extracted by correlating two detectors separated by a baseline whose length is comparable to the de Broglie wavelength.
We generalize the formalism to an array with multiple vector sensors, each of which is sensitive to a certain spatial component of the observables, as shown in Sec.\,\ref{VS}. 
The covariance matrix, which contains the information of the correlations between each pair of sensors, is expanded with additional $3 \times 3$ labels on the sensors' sensitive directions.

\subsection{Scalar Sensor Interferometry}

One starts with $\mathcal{N}_d$ detectors for scalar-like observables labeled with $I, J, \cdots$ and operates them simultaneously over time $T$. A discrete Fourier transform to the frequency space, indexed by an integer $k$, gives the expected response of the $I$th detector,
\begin{equation}
    \tilde{d}_I^k (\theta_{\rm sig}) = A_I (\omega_k) \, \tilde{a}\left(\omega_k, \vec{x}_I\right)+\tilde{n}_I \left( \omega_k \right), \label{monodata}
\end{equation}
where $A_I$ is the response functions of the detector that absorbs the coupling constant, $\tilde{a}$ comes from the transformation of the axion field (\ref{Sa}) to the frequency space, and $\tilde{n}_I$ is the Gaussian noise at frequency $\omega_k=2\pi k/T$. $\theta_{\rm sig}$ contains all the relevant signal parameters that characterize the axion contribution, for example, axion mass $m$ and the momentum distribution $f(\vec{p}\,)$ that we assume to be the same within the network of the detectors in this study. Since Eq.\,(\ref{monodata}) is a Gaussian random variable with zero mean, one can construct the multi-detector covariance matrix
\begin{equation}
    \Sigma_k (\theta_{\rm sig})  = \cor{ \tilde{d}^{k} \tilde{d}^{k \dagger}}\label{CMk}
\end{equation}
in terms of the $\mathcal{N}_d$-dimensional vector
\begin{equation}
    \tilde{d}^k \equiv \left(\tilde{d}_1^k,\tilde{d}_2^k,\cdots,\tilde{d}_{\mathcal{N}_d}^k\right)^T,
\end{equation}
where $\langle\cdots\rangle$ denotes the ensemble average on the stochastic fields and the statistical average on the detector noise. We leave the discussion of the ensemble average on the stochastic fields to the Appendix.
Each component of the covariance matrix $\Sigma_k$ characterizes the correlation between the $I$th and the $J$th detector,
\begin{equation}
\begin{aligned}
  \cor{\tilde{d}_I^{k} \tilde{d}_J^{k*}}=\delta_{IJ} \lambda_I\left(\omega_k\right) + \frac{A_I \left(\omega_k\right) A_J \left(\omega_k\right) \rho }{\bar{\omega}} F_{IJ}\left(\omega_k,\vec{x}_{IJ}\right),
\end{aligned}
\end{equation}
where $\lambda_I=\cor{\tilde{n}_I\tilde{n}_I^*}$ is the noise power spectral density of the $I$th detector, and
\begin{equation}
    F_{IJ}(\omega_k,\vec{x}_{IJ})=\int \frac{d^3\vec{p}}{\omega} \, f(\vec{p}) \, e^{\mi\, \vec{p}\cdot \vec{x}_{IJ}}\, \delta\left(\omega-\omega_k\right).
    \label{monof}
\end{equation}
represent the correlations of the axion field at the two locations separated by $\vec{x}_{IJ} \equiv \vec{x}_I-\vec{x}_J$, containing all the relevant signal parameters $\theta_{\rm sig}$ in this study.

With the help of the covariance matrix, we next construct the likelihood for an array of detectors in terms of data $\tilde{D}$ and the putative signal parameters $\theta_{\rm sig}$ \cite{Foster:2017hbq,Foster:2020fln},
\begin{equation}
    \mathcal{L}\left(\tilde{D}|\theta_{\rm sig}\right) = \prod_{k} \frac{\exp\left[-\frac{1}{2}\, \tilde{D}^{k\dagger} \cdot \Sigma_k^{-1} (\theta_{\rm sig}) \cdot \tilde{D}^k\right]}{\sqrt{|2\pi \Sigma_k (\theta_{\rm sig})|}},
\end{equation}
as well as the test statistic (TS)
\begin{equation}
    \Theta=2\left[\, \ln \,\mathcal{L}\left(\tilde{D}|\theta_{\rm sig}\right)-\ln \, \mathcal{L}\left(\tilde{D}|\theta_{\rm sig}=0\right) \,\right]\label{TS}
\end{equation}
that quantifies the significance of $\theta_{\rm sig}$.

To get an intuitive understanding of how to extract information from the TS in Eq.\,(\ref{TS}), we use the Asimov data set \cite{Cowan:2010js} where the average on the covariance of data $\tilde{D}$ converges to the truth value
\begin{equation}
\begin{aligned}
  &\cor{{\tilde{D}}_I^{k} {\tilde{D}}_J^{k*}} = \Sigma_k ( \theta_{\rm truth}) \\
  = \,&\delta_{IJ} \lambda_I\left(\omega_k\right)+ \frac{A_I \left(\omega_k\right) A_J \left(\omega_k\right) \rho}{\bar{\omega}} F^t_{IJ}\left(\omega_k,\vec{x}_{IJ}\right),
\end{aligned}
\end{equation}
where we use a superscript $t$ to label functions containing the truth parameters within $\theta_{\rm truth}$.
Under the assumptions that the noise is much larger than the signal, and the frequency bin size $2\pi/T$ is small enough compared to the varying scales of both the signals and the noise, we take the expectation value of the likelihood in Eq.\,(\ref{TS}) with the average data and get the asymptotic TS \cite{Foster:2020fln},
\begin{equation}
\begin{aligned}
    \widetilde{\Theta} (\theta_{\rm sig}) \approx \sum_{I,J}\frac{ \rho^2 T}{2 \pi \bar{\omega}^2} & \int d\omega\, \frac{A_I^2(\omega) A_J^2(\omega)}{\lambda_I(\omega) \lambda_J(\omega)} \, F_{JI}(\omega,\vec{x}_{JI})\\
    &\left( F^t_{IJ}(\omega,\vec{x}_{IJ})-\frac{1}{2}F_{IJ}(\omega,\vec{x}_{IJ})\right).
    \label{theta tilde}
\end{aligned}
\end{equation}
The maximization of the asymptotic TS (\ref{theta tilde}) happens when $\theta_{\rm sig}$ in $F_{IJ}$ approaches $\theta_{\rm truth}$. 

One can further estimate the Fisher information matrix in terms of different parameters $\theta_i \in \theta_{\rm sig}$, evaluated at the truth parameters
\begin{equation}
    I_{ij} (\theta_{\rm sig}) = -\frac{1}{2} \frac{\partial^2 \widetilde{\Theta}}{\partial \theta_i \, \partial \theta_j} |_{\theta_{\rm sig}=\theta_{\rm truth}}.
    \label{var est}
\end{equation}
For single parameter $\theta_i$, the inverse of Eq.\,(\ref{var est}) gives the uncertainty $\sigma^2_{\theta_i}$.

Take an infinitely cold stream with $f (\vec{p}\,) = \delta^3 (\vec{p} - \vec{p}_0)$ as an example. With two detectors separated by a baseline $\vec{x}_{IJ}$, one can extract the information $\cos \theta_{\rm str} \equiv  \hat{p}_0 \cdot \hat{x}_{IJ}$ from the relative phase factor in Eq.\,(\ref{monof}), with the uncertainty $\sigma^2_{\theta_{\rm str}} \propto 1/\left( p_0\,x_{IJ} \, \sin \theta_{\rm str} \right)^2$ derived from Eq.\,(\ref{var est}) \cite{Foster:2020fln}. 

In the case in which the baselines $\vec{x}_{IJ}$ between each pair of axion detectors cover a large volume in the coordinate space with a sufficient time resolution, we can take the continuous limit $\omega_k\rightarrow\omega$ and $\vec{x}_{IJ}\rightarrow \vec{x}$, which leads to $F_{IJ} (\omega_k,\vec{x}_{IJ})\rightarrow F (\omega, \vec{x})$. 
Thus the three-dimensional Fourier transform
\begin{equation}
 \begin{aligned}
   \int \frac{d^3 \vec{x}}{(2\pi)^3} \, \me^{-\mi\, \vec{p}\cdot \vec{x}} \, \int d \omega \, F(\omega,\vec{x}) = \frac{f(\vec{p}\,)}{\omega}
\end{aligned}
\label{mono iFT}
\end{equation}
enables us to recover the complete information of the axion field in momentum space, namely, $f(\vec{p}\,)$.

\subsection{Vector Sensor Array}

In this subsection, we consider a network of vector sensors. For each detector, we replace the expected response in Eq.\,(\ref{monodata}) with the vector-like signals $\vec{\mathcal{O}} \left(\omega_k,\vec{x}_I\right)$,
\begin{equation}
    \tilde{d}_I^k=A_I(\omega_k) \,  \vec{\mathcal{O}} \left(\omega_k,\vec{x}_I\right) \cdot \hat{l}_I+\tilde{n}_I \left(\omega_k\right),
\end{equation}
where $\hat{l}_I$ is a unit vector representing the sensitive direction of the sensor. Similar to Eq.\,(\ref{monof}), the correlations of the vector-like signals are
\begin{widetext}
\begin{equation}
    F_{IJ} \left( \omega_k, \vec{x}_{IJ}, \hat{l}_I,\hat{l}_J \right) = \int \frac{d^3\vec{p}}{\omega}\, f(\vec{p}\,)\, \me^{\mi\, \vec{p} \cdot \vec{x}_{IJ}}\, \delta\left(\omega-\omega_k\right)\, \zeta^2_\mathcal{O} (p) \, \sum_{\beta}h(\beta)\, \left(\vec{\epsilon}_{\beta}\cdot \hat{l}_I\right)\, \left(\vec{\epsilon}_{\beta}^{\,*}\cdot \hat{l}_J\right),
    \label{dipolef}
\end{equation}
\end{widetext}
 where $\zeta_\mathcal{O} (p)$ is the factor before $\Phi_j$ in  Eq.\,(\ref{OAG} - \ref{OEDM}), depending on different types of coupling $\mathcal{O}$, and $h(\beta)$ is the percentage of the polarization mode $\beta$. 
Compared with the case of the scalar sensors in Eq.\,(\ref{monof}), the correlations of the vector-like signals (\ref{dipolef}) have additional polarization-dependent factors $h(\beta)\, \left(\vec{\epsilon}_{\beta}\cdot \hat{l}_I\right)\left(\vec{\epsilon}_{\beta}^{\,*}\cdot \hat{l}_J\right)$. Thus the product $f(\vec{p}\,) \times \zeta^2_\mathcal{O} (p)$ can be extracted only when we know $h(\beta)$. 

We next rewrite Eq.\,(\ref{dipolef}) as
\begin{equation}
\begin{aligned}
    &F_{IJ}(\omega_k, \vec{x}_{IJ}, \hat{l}_I,\hat{l}_J )\\
    =\,& p_k\, \zeta^2_\mathcal{O} (p_k) \, \sum_{\beta} h(\beta) \,\mathcal{F}^{\beta}_{IJ} (p_k,\vec{x}_{IJ},\hat{l}_I,\hat{l}_J),\label{dipf2}
\end{aligned}
\end{equation}
where $p_k = \sqrt{\omega_k^2-m^2}$ and
\begin{equation}
\begin{aligned}
    &\mathcal{F}^{\beta}_{IJ}(p_k,\vec{x}_{IJ},\hat{l}_I,\hat{l}_J) \\
    \equiv \, & \int d^2  \hat{\Omega} \, f(p_k,\hat{\Omega})\, \me^{\mi\, p_k\, \hat{\Omega} \cdot \vec{x}_{IJ}}\,\left(\vec{\epsilon}_{\beta}\cdot \hat{l}_I\right)\left(\vec{\epsilon}_{\beta}^{\,*}\cdot \hat{l}_J\right),\label{fbeta}
\end{aligned}
\end{equation}
which characterizes the correlations between the two vector-like signals with a specific polarization mode $\beta$ and energy $\omega_k$.

\begin{figure}[htb]
    \centering
    \includegraphics[width=0.8\columnwidth]{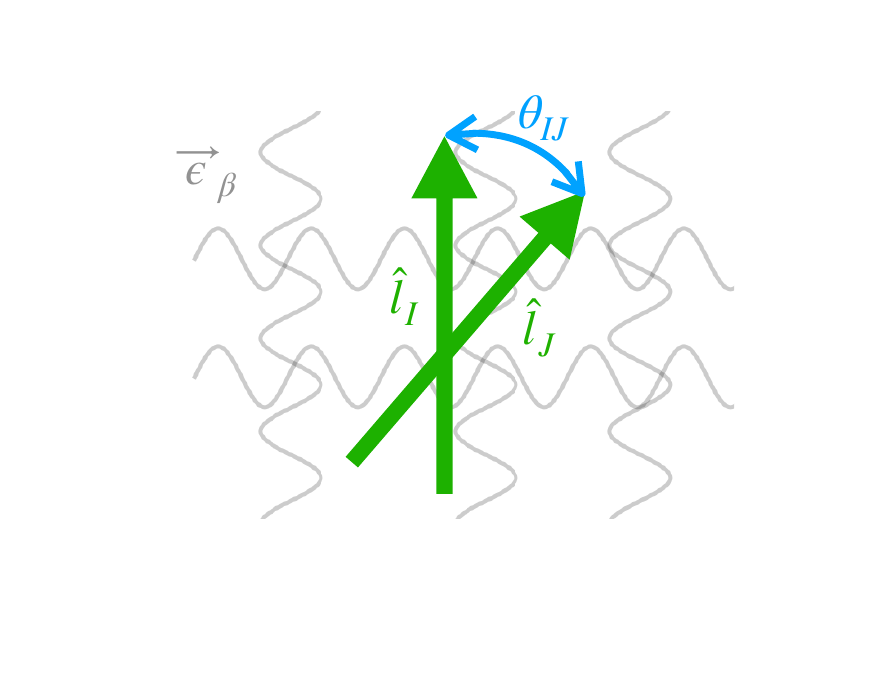}
    \caption{{A schematic diagram for a pair of vector sensors distributed at the same location with different sensitive directions $\hat{l}_I$ and $\hat{l}_J$ is shown.  The green arrows indicate the sensitive directions of the vector sensors, and the gray wavy lines represent the stochastic bosonic fields with wave functions $\vec{\epsilon}_\beta$. 
   If the source is isotropic, the angular correlations always show the universal dipole correlation which is proportional to $\hat{l}_I\cdot \hat{l}_J = \cos \theta_{IJ}$, according to Eq.\,(\ref{iso overlap}). }}
    \label{2dip}
\end{figure}

 Here we show an example of what information can be extracted if we tune $\hat{l}_I$. We can put two detectors with different sensitive directions, $\hat{l}_I$ and $\hat{l}_J$, at the same location. The schematic diagram of the detectors is shown in Fig.\,\ref{2dip}.
For a background source with an isotropic momentum distribution
\be f_{\rm iso}(p,\hat{\Omega})= \frac{f_{\rm iso}(p)}{4\pi p^2},\label{PDFI}\ee
where $p=|\vec{p}\,|$, the signal correlations between the longitude modes are
\begin{equation}
\begin{aligned}
    \mathcal{F}^{\,0}_{IJ} \left(p,\hat{l}_I,\hat{l}_J\right) =& \int d^2 \hat{\Omega}\, \frac{f_{\rm iso} (p)}{4 \pi p^2}\, \left( \vec{\epsilon}_0 \cdot \hat{l}_I \right)\left( \vec{\epsilon}_0 \cdot \hat{l}_J\right)\\
    =&\frac{\omega^2 f_{\rm iso} (p)}{3 m^2 p^2} \, \hat{l}_I \cdot \hat{l}_J,
\end{aligned}
\end{equation}
and the ones for each type of transverse modes are
\begin{equation}
\begin{aligned}
    \mathcal{F}^{R/L}_{IJ} \left(p,\hat{l}_I,\hat{l}_J\right) =& \int d^2 \hat{\Omega}\,  \frac{f_{\rm iso} (p)}{4\pi p^2}\, \left(\vec{\epsilon}_{R/L}\cdot \hat{l}_I\right)\left(\vec{\epsilon}_{R/L}^{\,*}\cdot \hat{l}_J\right)\\
    =&\frac{f_{\rm iso} (p)}{3 p^2}\, \hat{l}_I \cdot \hat{l}_J.
    \label{isoLR}
\end{aligned}
\end{equation}
Among all the polarization modes, the angular correlations are universally proportional to the dipole correlation \cite{Jenet:2014bea}
\begin{equation}
    \mathcal{F}^{\beta}_{IJ} \propto \hat{l}_I \cdot \hat{l}_J=\cos \theta_{IJ}, \label{iso overlap}
\end{equation}
where $\theta_{IJ}$ is the angle between $\hat{l}_I$ and $\hat{l}_J$. In the presence of many vector sensors with different sensitive directions, such as the setup proposed in \cite{Smiga:2021acs}, any deviation from Eq.\,(\ref{iso overlap}) is a sign of anisotropy for the sources. Note that the converse is not necessarily true. As we will see in Sec.\,\ref{IDM}, for non-relativistic sources with all the polarization modes contributing equally, anisotropic distribution can also lead to Eq.\,(\ref{iso overlap}).

In the following sections, we will study further the implications of Eq.\,(\ref{dipolef}), and how we can arrange  $\hat{l}_I$ and $\vec{x}_{IJ}$ to investigate the macroscopic properties and microscopic nature of the vector-like sources.

\section{Identification of the Couplings}\label{S4}
In this section, we study the correlations of the vector-like signals (\ref{dipf2}) with distinctive momentum distributions and polarization modes, with which one can identify the properties of the bosonic background.

\subsection{Source from a Specific Direction}\label{SFD}

The first case we consider is the source from a specific direction $\hat{\Omega}_0$, e.g. the cold stream \cite{OHare:2017yze, Foster:2017hbq, Knirck:2018knd}, whose momentum distribution is
\be f_{\rm str}(p,\hat{\Omega})= \frac{f_{\rm str}(p)}{p^2}\, \delta^2(\hat{\Omega}-\hat{\Omega}_0). \label{str PDF}\ee
 Take the above expression into Eq.\,(\ref{fbeta}) and choose $x_{IJ} = 0$, the longitudinal-mode-only case gives
\begin{equation}
    \mathcal{F}^{\,0}_{IJ} \left(p,\hat{l}_I,\hat{l}_J\right) = \frac{\omega^2 f_{\rm str}(p)}{m^2 p^2} \left( \hat{\Omega}_0 \cdot \hat{l}_I \right)\left( \hat{\Omega}_0 \cdot \hat{l}_J \right).\label{FFD}
\end{equation}
The maximization of Eq.\,(\ref{FFD}) happens when both $\hat{l}_I$ and $\hat{l}_J$ are parallel with $\hat{\Omega}_0$. For transverse modes without macroscopic polarization, Eq.\,(\ref{fbeta}) becomes
\begin{equation}
\begin{aligned}
    &\mathcal{F}^R_{IJ} \left(p,\hat{l}_I,\hat{l}_J\right) + \mathcal{F}^L_{IJ} \left(p,\hat{l}_I,\hat{l}_J\right) \\
    =\,& \frac{f_{\rm str}(p)}{p^2} \left( \hat{l}_I \cdot \hat{l}_J - ( \hat{\Omega}_0 \cdot \hat{l}_I)\,( \hat{\Omega}_0 \cdot \hat{l}_J )  \right),
    \label{FRFL}
\end{aligned}
\end{equation}
which vanishes when the three directions are parallel with each other and reaches the maximum when $\hat{l}_I$ and $\hat{l}_J$ take the same direction on the transverse plane of $\hat{\Omega}_0$.
We will discuss the localization of such signals in detail in Sec.\,\ref{SL}.

\subsection{Isotropic Cosmological Background}

\begin{figure*}[htb] 
\includegraphics[width=0.8\columnwidth]{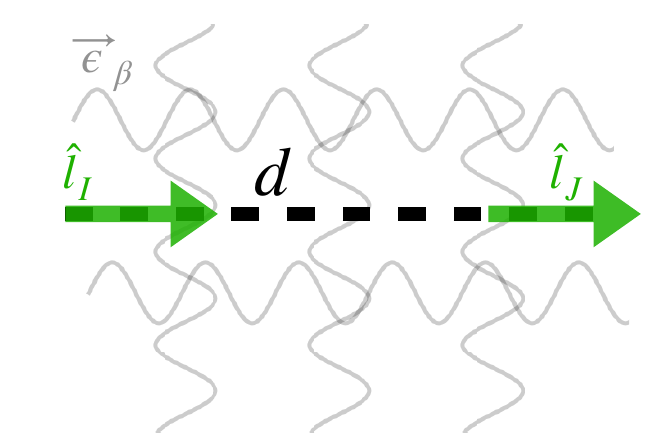}\qquad \quad
\includegraphics[width=0.8\columnwidth]{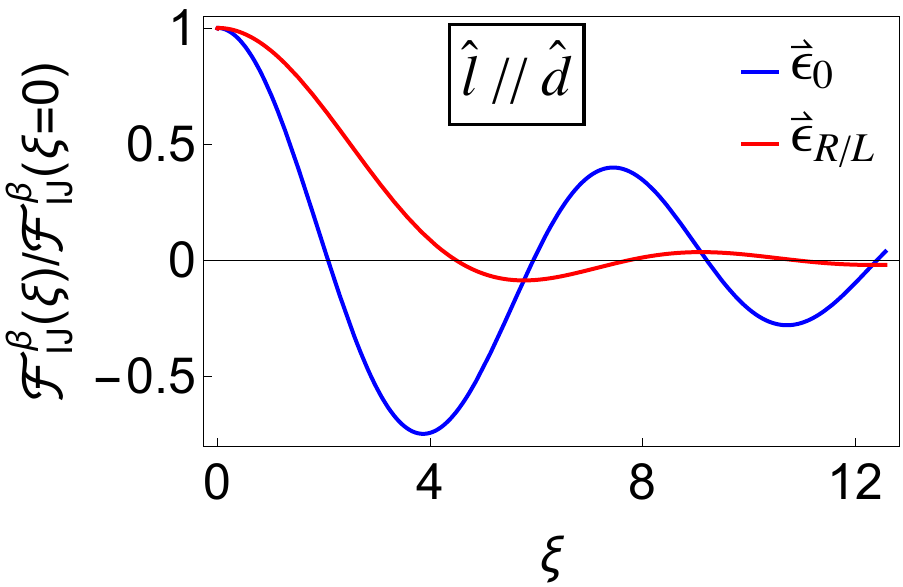}\\
\includegraphics[width=0.8\columnwidth]{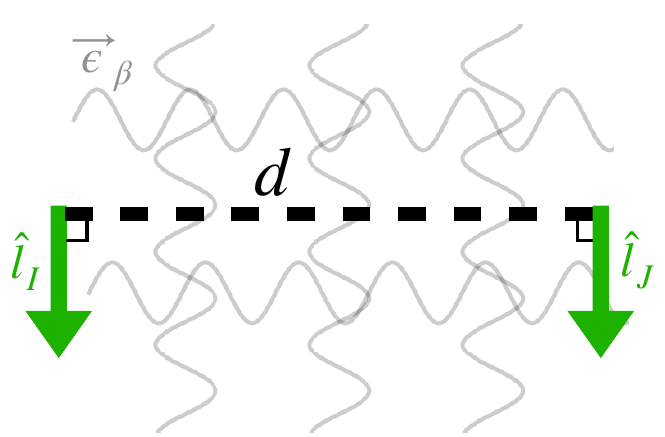}\qquad \quad
\includegraphics[width=0.8\columnwidth]{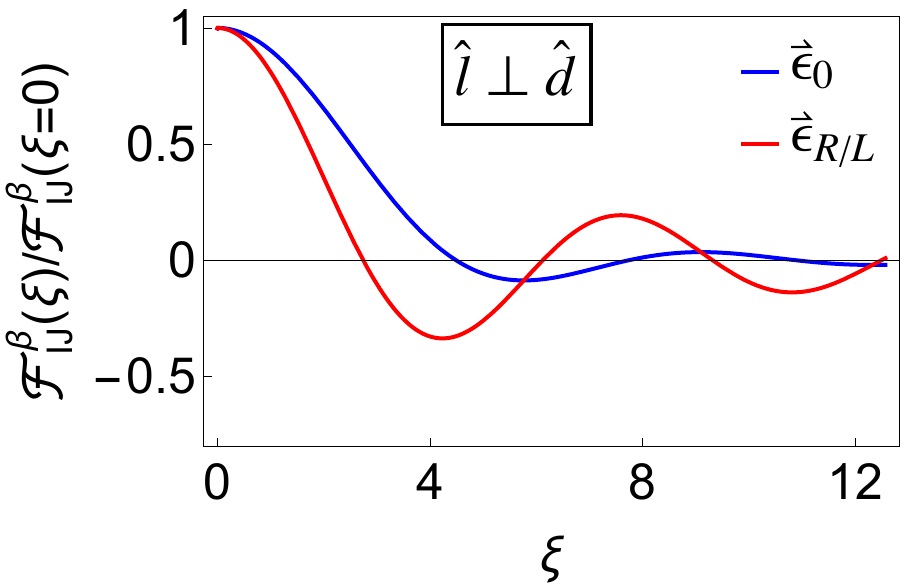}
\caption{{The left panels show two schematic diagrams for different arrangements of the vector sensors, shown as the green arrows, separated by the baseline $d$, represented by the black dashed lines. The sensitive directions are parallel with $\hat{d}$ in the upper panels while perpendicular to it in the lower ones. The gray wavy lines are the isotropic bosonic fields with either longitudinal or transverse degrees of freedom.  The corresponding correlations in Eq.\,(\ref{ISOPP} - \ref{ISOTT}) are shown in the right  panels. One can see that, at $\xi \equiv p d \approx 4$, $\epsilon_0$ and $\epsilon_{L/R}$ show distinct correlations.}} 
\label{fig:fpara}
\end{figure*}

We next consider the isotropic bosonic field background with the momentum distribution in Eq.\,(\ref{PDFI}), which usually comes from a cosmological origin. As shown in Eq.\,(\ref{iso overlap}), when two vector sensors are at the same location, they demonstrate the universal dipole correlation, independent of the polarization modes of the source. We will show that by separating two detectors by a baseline with a finite length $d = x_{IJ}$, we are able to distinguish the longitudinal modes of the source from the transverse ones. Note that, because $d$ always comes together with momentum $p$ in Eq.\,(\ref{fbeta}), we introduce the dimensionless parameter $\xi \equiv p \, d$.

We consider the setup with $\hat{l}_I \parallel \hat{l}_J$ in this subsection, where the pair of vector sensors respond to right-hand and left-hand circular polarization modes identically. 
When both of the sensitive directions are parallel with $\hat{d}$, Eq.\,(\ref{fbeta}) gives
\begin{gather}
    \mathcal{F}^{\,0}_{IJ} \left(p,\xi,\hat{l}_I \parallel \hat{l}_J \parallel \hat{d}\right) =\frac{\omega^2 f_{\rm iso} (p)}{m^2 p^2}\frac{2\xi \cos \xi+\left( \xi^2-2\right)\sin \xi }{\xi^3}, \label{ISOPP}  \\
    \mathcal{F}^{R/L}_{IJ}\left(p,\xi,\hat{l}_I  \parallel \hat{l}_J \parallel \hat{d}\right) = \frac{f_{\rm iso} (p)}{p^2}\, \frac{\sin \xi-\xi \cos \xi}{\xi^3}, \label{ISOPT}
\end{gather}
the results of which are shown in the upper panels of Fig.\,\ref{fig:fpara}.

Similarly, when the sensitive directions are perpendicular to $\hat{d}$ and parallel with each other, we have
\begin{gather}
    \mathcal{F}^{\,0}_{IJ} \left(p,\xi,\hat{l}_I \parallel \hat{l}_J \perp \hat{d}\right) = \frac{\omega^2 f_{\rm iso} (p)}{m^2 p^2} \, \frac{\sin \xi-\xi \cos \xi}{\xi^3}, \label{ISOTP}\\
    \mathcal{F}^{R/L}_{IJ} \left(p,\xi,\hat{l}_I \parallel \hat{l}_J \perp \hat{d}\right) = \frac{f_{\rm iso} (p)}{p^2} \, \frac{\xi \cos \xi+\left( \xi^2-1\right)\sin \xi}{2\xi^3}, \label{ISOTT}
\end{gather}
which is shown in the lower panels of Fig.\,\ref{fig:fpara}. Comparing the two different ways to arrange the sensitive directions, one can see that significant differences appear at $\xi \approx 4$. The longitudinal mode correlation vanishes, and transverse modes correlations reach the minimum for the perpendicular setup, while the features flip for the parallel one.

We will see in Sec.\,\ref{CP} that a pair of vector sensors with non-parallel sensitive directions and a long baseline can further distinguish between the right-hand and left-hand circular polarization modes.

\subsection{Standard Halo Model of Dark Matter}\label{IDM}

The last example we consider is the Maxwellian distribution of the standard halo model of cold dark matter, whose velocity distribution is \cite{Derevianko:2016vpm}
\be f_{\rm DM} (\vec{v}\,)= \left(2\pi \vvir^2\right)^{-3/2} \, \exp \left[- \frac{(\vec{v}-\vec{v}_g)^2}{2 \vvir^2} \right], \label{fDM}
\ee
where $\vvir\approx 10^{-3}$ and $\vec{v}_g$ are the virial velocity and the Earth's velocity in the galactic reference frame respectively. In the non-relativistic limit, Eq.\,(\ref{dipolef}) can be approximated as
 \begin{equation}
 \begin{aligned}
     F_{IJ} \left( \omega_k, \vec{d}, \hat{l}_I,\hat{l}_J \right) &\approx \int  \frac{d^3 \vec{v}}{m^2\, v} \, f_{\rm DM}(\vec{v})\, \me^{\mi \, m \, \vec{v}\cdot \vec{d}}\, \delta \left( v-v_k \right)  \\
     \times &  \zeta^2_\mathcal{O} (m\, v) \, \sum_{\beta}h(\beta)\, \left(\vec{\epsilon}_{\beta}\cdot  \hat{l}_I\right)\, \left(\vec{\epsilon}_{\beta}^{\,*}\cdot \hat{l}_J\right) ,\label{FNR}
\end{aligned}
 \end{equation}
 where $v_k$ satisfies $\omega_k/m = 1 + v_k^2/2$.  $\vec{\epsilon}_\beta$ reduce to unit directional vectors and $h(\beta)$ contains an equal percentage among the allowed polarization modes in  Eq.\,(\ref{OAG} - \ref{OEDM}). With a preknowledge of the velocity distribution in Eq.\,(\ref{fDM}), one can break the degeneracy in the product $f_{\rm DM}(\vec{v}) \times \zeta^2_\mathcal{O} (m\, v)$ and identify the exact type of the coupling $\mathcal{O}$ listed in Sec.\,\ref{VS}.

Due to the complicated forms of Eq.\,(\ref{FNR}) defined at each energy bin and the narrow bandwidth of the standard halo dark matter, we define the dark matter correlation matrix as
\begin{equation} 
\begin{aligned} 
&\Big(\mathscr{C}_\mathcal{O} ( \vec{d}\, ) \Big)_{ij}\\
      \equiv& \int  dv_k\, v_k\, \re\left[F_{IJ} ( \omega_k, \vec{d}, \hat{i}, \hat{j} )\right]\\
     \approx& \int   \frac{d^3 \vec{v}}{m^2} \, f_{\rm DM}(\vec{v}\,)\, \cos \left[ m \, \vec{v}\cdot \vec{d}\, \right]\, \zeta^2_\mathcal{O} (m\, v) \, \sum_{\beta}h(\beta)\, \epsilon_{\beta}^i\, \epsilon_{\beta}^{j} ,
\end{aligned}
 \label{g_ij}
\end{equation}
which comes from the integration of the real part of Eq.\,(\ref{FNR}) in energy. Eq.\,(\ref{g_ij}) always leads to analytic expressions from the Gaussian integral.
$i$ and $j$ denote the index of the spatial components in Cartesian coordinates. For dark matter, we choose the linear polarization basis where $\vec{\epsilon}_\beta$ are real.
Equation ({\ref{g_ij}}) is the vectorial generalization of the two-point spatial correlation function of the scalar-like signals from the dark matter in \cite{Derevianko:2016vpm} and the Appendix. The daily modulations of the vector-like signals were studied in \cite{Lisanti:2021vij,Gramolin:2021mqv} for the axion gradient and in \cite{Caputo:2021eaa} for the kinetic-mixing dark photon.

\subsubsection{Axion Gradient}

For the axion gradient, there is only longitudinal mode and $\zeta_a = m\,v$. To give an intuitive understanding, we can decompose Eq.\,(\ref{g_ij}) into three parts:
\begin{equation}
    \mathscr{C}_a(\vec{d}\,)= \,\exp \left[-\frac{d^2}{2 \lambda_c^2}\right]\, \left(C_a^{\vir}+C_a^{g}+C_a^{g,\vir}\right),
    \label{axiondecomp}
\end{equation}
where  $C_a^{\vir}$ contains a term proportional to the square of the virial velocity $v_{\rm vir}$, $C_a^{g}$ comes from the square of $v_g$, and $C_a^{g,\vir}$ is a product of the two. 
$\lambda_c \equiv 1/\left(m\, \vvir\right)$ is the correlation length of the virial dark matter.
The overall exponential factor indicates that the correlation decays at a length scale $\lambda_c$ due to the virial fluctuations.

Without loss of generality, we assume that the separation between the detectors pair $\vec{d}$ is along the $\hat{z}$-axis, where the virial part takes the form
\begin{equation}
    C_a^{\vir}=\vvir^2\,\diag\left(1,1,1-\frac{d^2}{\lambda_c^2}\right)\,\cos\left[m\,\vec{v}_g \cdot \vec{d}\, \right]. \label{iso}
\end{equation}
In the absence of $\vec{v}_g$, 
Eq.\,(\ref{iso}) is the only non-vanishing one. The anisotropic velocity leads to a spatial oscillation factor $m\,\vec{v}_g \cdot \vec{d}$. The second part of Eq.\,(\ref{axiondecomp}) is
\begin{equation}
    \Big(C_a^{g}\Big)_{ij}=v_g^i\, v_g^j  \cos\left[m\,\vec{v}_g\cdot \vec{d}\,\right], \label{aniso}
\end{equation} 
which is an anisotropic term induced by $\vec{v}_g$. The last term in  Eq.\,(\ref{axiondecomp}) is
\begin{equation} 
\begin{aligned}
C_a^{g,\vir}= \vvir\,
\begin{pmatrix}
    0 & 0 & -v_g^x  \\
    0 & 0 & -v_g^y  \\
    -v_g^x  & -v_g^y  & -2v_g^z
\end{pmatrix} \, 
\frac{d}{\lambda_c}\, \sin\left[m\,\vec{v}_g \cdot \vec{d}\,\right],
\end{aligned}\label{cross}
\end{equation}
which describes the mixing effect between the anisotropic velocity $\vec{v}_g$ and the virial one.

\begin{figure*}[htb]
    \centering
    \includegraphics[width=0.9\columnwidth]{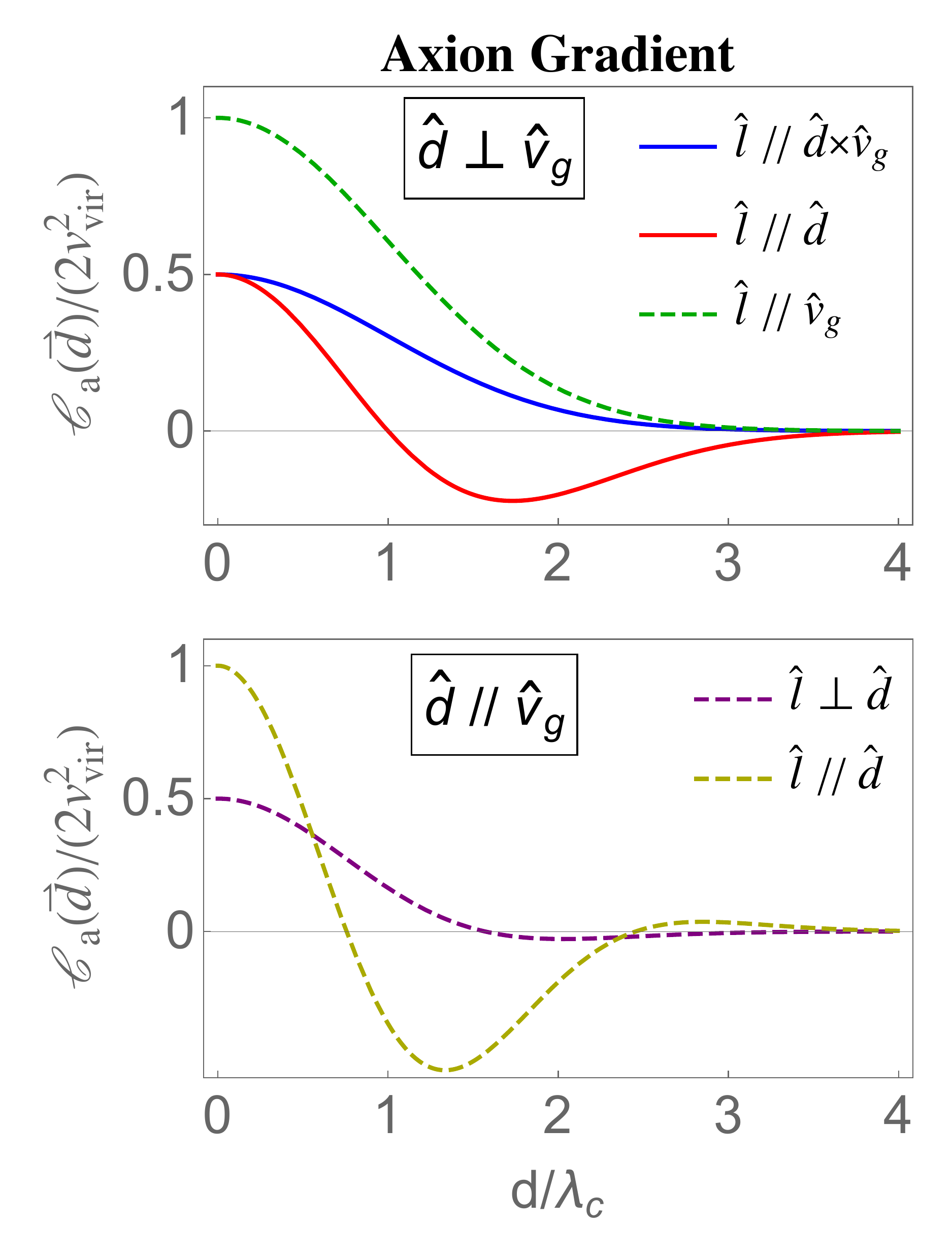}\quad
    \includegraphics[width=0.9\columnwidth]{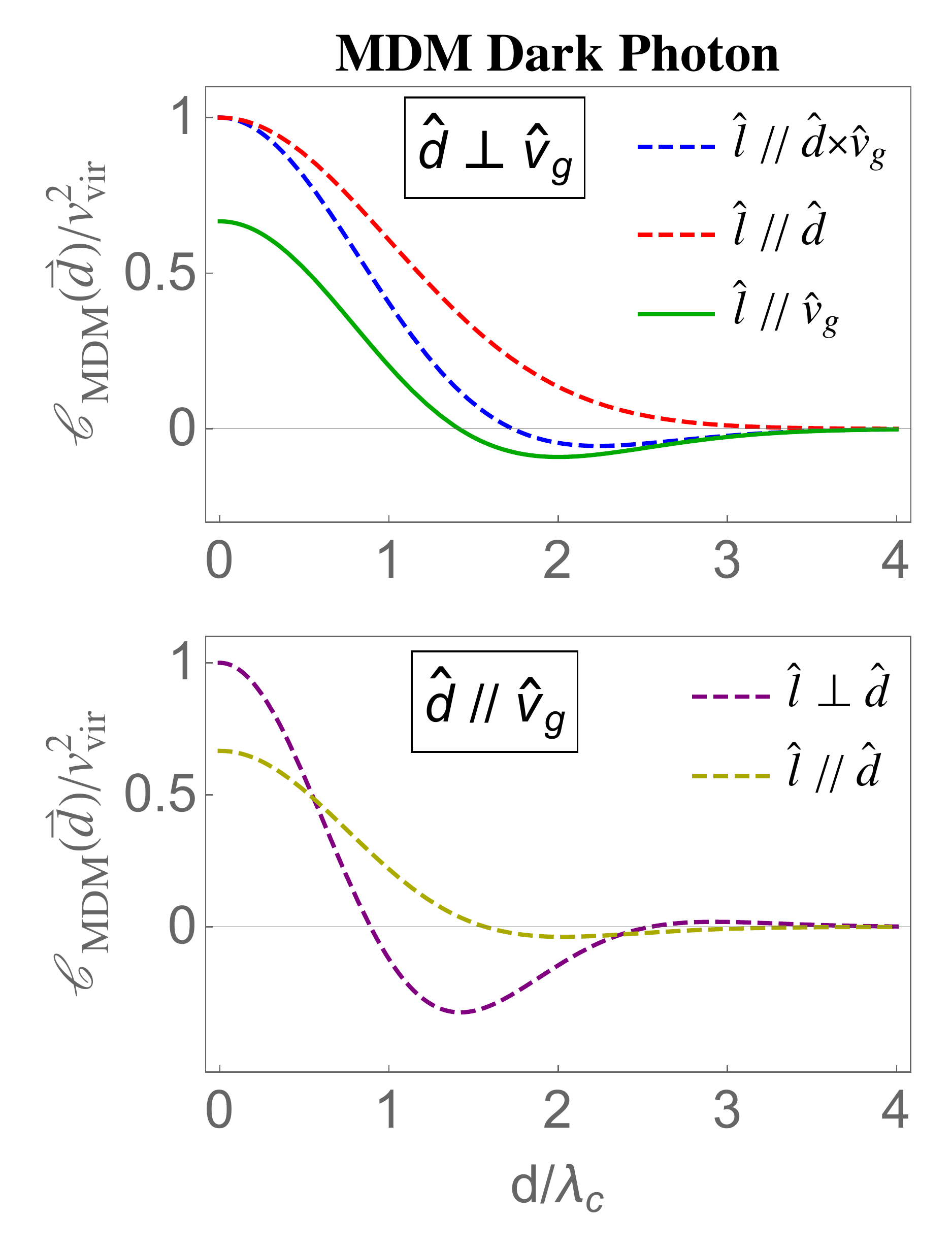}
    \caption{ Two-point spatial correlation functions for the axion gradient in Eq.\,(\ref{axiondecomp}) with only the longitudinal modes and the MDM dark photon in Eq.\,(\ref{CMDM}) with the transverse interacting modes, are shown in the left and right panels, respectively. The two sensitive directions of the detectors are assumed to be coaligned such that there are five different ways to arrange the directions of $\hat{d}$ and $\hat{l}$ in terms of $\hat{v}_g$ in each case.
    The upper and lower panels include the correlations when $\hat{d}$ is perpendicular and parallel with $\hat{v}_g$ respectively. 
    %We use solid lines to label the case without the contribution from $\hat{v}_g$ while the dashed lines show the cases that are influenced by it.
    We use solid lines to label the cases independent of the norm of $\vec{v}_g$, i.e., $v_g$, while the dashed lines show the cases influenced by it, see discussion after Eq.\,(\ref{cross}).
 } 
    \label{fig:wind}
\end{figure*}

Different cases of $\mathscr{C}_a(\vec{d}\,)$ are shown in the left panels of Fig.\,\ref{fig:wind}. For simplicity, we only consider the situation in which two detectors are co-aligned. In such a case, there are five different ways to arrange the directions of the baseline $\hat{d}$ and the sensitive directions $\hat{l}$. 
$\vvir=v_g$ is assumed in this work. When $\hat{v}_g \perp \hat{d}$, both the spatial oscillation factor $m\,\vec{v}_g\cdot \vec{d}$ and the mixing term  (\ref{cross}) vanish. If one further tunes $\hat{l} \perp \hat{v}_g$,  Eq.\,(\ref{aniso})  vanishes as well.
Then $\mathscr{C}_a(\vec{d}\,)$ is influenced by the virial velocity only, so that it decreases to zero monotonically in terms of $d$.
On the other hand, when $\hat{l}\, \parallel\, \hat{d}$, $\mathscr{C}_a(\vec{d}\, )$ becomes negative first before  decaying to zero.
The impact of $\vec{v}_g$ shows up in the rest cases where the spatial oscillation factor $m\,\vec{v}_g\cdot \vec{d}$ can appear. However, since the correlation decays significantly  when $d \simeq 2\pi/\left(m\, v_g\right)$, we can hardly see this oscillation.

Notice that if one defines Eq.\,(\ref{g_ij}) using the imaginary part of Eq.\,(\ref{FNR}), the results for Eq.\,(\ref{iso}-\ref{cross}) simply differ by a $\pi/2$ phase delay. Thus Eq.\,(\ref{g_ij}) already contains all the information that can be extracted.

We also calculate the angular correlations between two detectors at the same location with different sensitive directions $\hat{l}_I$ and $\hat{l}_J$, which are the projection of $\mathscr{C}_a(0)$ on these two directions,
\begin{equation}
\begin{aligned}
      &\Gamma_a\left(\hat{l}_I,\hat{l}_J\right)\\
      \equiv&\left(\hat{l}_I\right)^{\textrm{T}} \cdot \mathscr{C}_a(0)\cdot \hat{l}_J\\
     =& \vvir^2 \,\hat{l}_I \cdot \hat{l}_J + \left(\vec{v}_g \cdot \hat{l}_I\right) \left(\vec{v}_g \cdot \hat{l}_J\right).
\end{aligned}
    \label{windhd}
\end{equation}
   The first term, which comes from the vector-like property of the signal, is again the universal dipole correlation shown in Eq.\,(\ref{iso overlap}),
   while the second term in Eq.\,(\ref{windhd}) describes the anisotropy introduced from  $\vec{v}_g$.

\subsubsection{Dark Photon}

For non-relativistic dark photons, the polarization vectors are equally distributed on the unit sphere in the galaxy frame. In the linear polarization basis, this is equivalent to taking the percentage of each polarization mode $\vec{\epsilon}_\beta$, i.e., $h(\beta)$, to be $1/3$. However, for specific types of interaction, some modes may not contribute.

We first consider the MDM dark photon whose transverse modes interact with the sensors and $\zeta_\MDM = m\,v$.
The right panels of Fig.\,\ref{fig:wind} show different cases of the MDM dark photon's two-point correlation functions $\mathscr{C}_\MDM$, which again can be decomposed into
\begin{equation}
    \mathscr{C}_\MDM(\vec{d}\,)=\frac{1}{3}\,\exp \left[-\frac{d^2}{2 \lambda_c^2}\right] \left(C_{\MDM}^{\vir}+C_{\MDM}^{g}+C_{\MDM}^{g,\vir}\right).\label{CMDM}
\end{equation}
The virial part is
\begin{equation}
    \begin{aligned}
         C_{\MDM}^{\vir}=\vvir^2\,\diag\left(2-\frac{d^2}{\lambda_c^2},2-\frac{d^2}{\lambda_c^2},2\right)\, \cos\left[m\,\vec{v}_g \cdot \vec{d}\,\right].
    \end{aligned} \label{mdmvir}
\end{equation}
Compared with Eq.\,(\ref{iso}), we can find the relation
\begin{equation}
    \Big(C_{\MDM}^{\vir}\Big)_{ii}=\sum_{n\neq i} \Big(C_{a}^{\vir}\Big)_{nn},
\end{equation}
since the directions of the axion gradient signal are perpendicular to the ones of the MDM dark photon with the same momentum. The second part of Eq.\,(\ref{CMDM}) takes the form
\begin{equation}
    \Big(C_{\MDM}^{g}\Big)_{ij}=\Big(v_g^2\, \delta_{ij}-v_g^i\, v_g^j\Big) \cos\left[m\,\vec{v}_g \cdot \vec{d}\,\right].
    \label{mdmg}
\end{equation}
In addition to the anisotropic term with a different sign compared with the axion gradient in Eq.\,(\ref{aniso}), there is also a diagonal part proportional to $v_g^2$. The mixing term is
\begin{equation}
    \begin{aligned}
         C_{\MDM}^{g,\vir}= \vvir\,
         \begin{pmatrix}
            -2v_g^z & 0 & v_g^x \\
            0 & -2v_g^z & v_g^y \\
            v_g^x & v_g^y & 0
         \end{pmatrix}\, \frac{d}{\lambda_c}\, \sin\left[m\,\vec{v}_g\cdot \vec{d} \,\right].
    \end{aligned} \label{mdmmix}
\end{equation}

The angular correlations of the MDM dark photon are
\begin{equation}
     \Gamma_{\MDM}\left(\hat{l}_I,\hat{l}_J\right)= \vvir^2\, \left(\hat{l}_I \cdot \hat{l}_J \right) -\frac{1}{3} \left(\vec{v}_g \cdot \hat{l}_I\right) \left(\vec{v}_g \cdot \hat{l}_J\right),
\end{equation}
whose anisotropic part has a different sign compared to that of the axion gradient in Eq.\,(\ref{windhd}). 

\begin{figure}[htb]
    \centering
    \includegraphics[width=0.9\columnwidth]{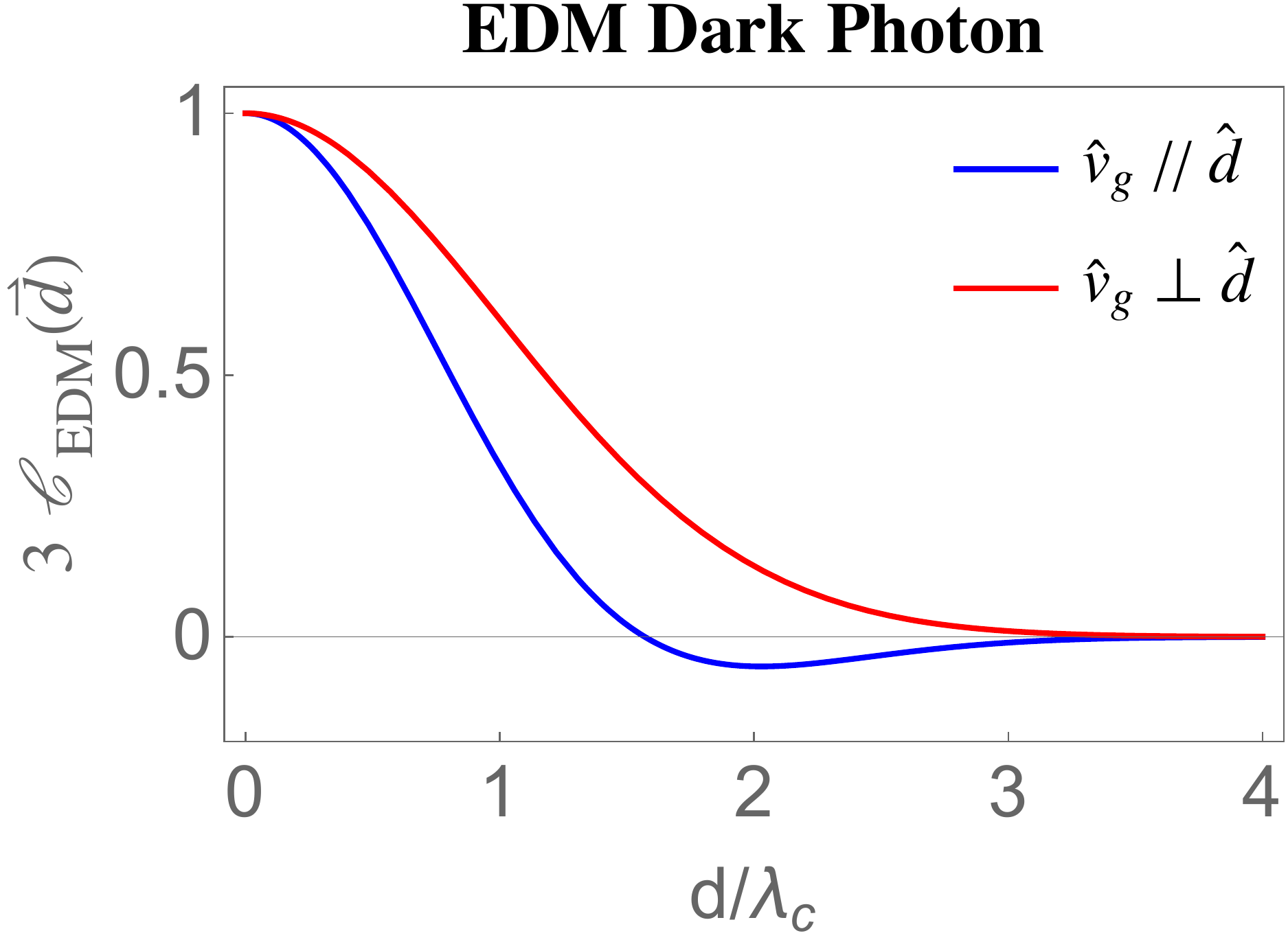}
    \caption{{The two-point spatial correlation functions for the EDM dark photon in Eq.\,(\ref{EDM cor}), with all the $3$ polarization modes contributing equally, is shown. }} 
    \label{fig:edm}
\end{figure}

Finally, we consider the EDM dark photon where $\zeta_\EDM = m$
and all three polarization modes contribute equally in the non-relativistic limit.
 Thus the correlation matrix (\ref{g_ij}) reduces to an identity insensitive to $\hat{l}$
\begin{equation}
    \mathscr{C}_\EDM(\vec{d}\,) =\frac{1}{3} \exp \left[-\frac{d^2}{2 \lambda_c^2}\right]\,\cos\left[m\,\vec{v}_g\cdot \vec{d}\,\right] \mathbb{I}_{3\times3},
    \label{EDM cor}
\end{equation}
which is shown in Fig.\,\ref{fig:edm}. This shares the same form as the scalar correlation in \cite{Derevianko:2016vpm} and the Appendix, with an additional factor of $1/3$ for each diagonal component.

The angular correlations of the EDM dark photon contain only the universal dipole correlation
\begin{equation}
    \Gamma_{\EDM}\left(\hat{l}_I,\hat{l}_J\right)=\frac{1}{3}\Big(\hat{l}_I \cdot \hat{l}_J\Big).\label{GEDM}
\end{equation}

As mentioned in Sec.\,\ref{NRDC}, for the axial-vector dark photon, the $U(1)_{B-L}/U(1)_{B}$ dark photon or the kinetic-mixing dark photon, the vector-like signals are proportional to their spatial components of the wave functions, the same as the EDM dark photon. Therefore, they share the same form of the correlations as the EDM dark photon in Eq.\,(\ref{EDM cor}) and (\ref{GEDM}), despite the different detector systems.
One can further distinguish between the EDM dark photon and the axial-vector by boosting the sensors to be relativistic, like the storage ring experiments \cite{Graham:2020kai} where the contribution from the longitudinal mode of the EDM dark photon is suppressed, as shown in Eq.\,(\ref{OEDM}).

In conclusion, one can distinguish between the different types of dark matter by the spatial and the angular correlations, based on the unique signals of each one discussed in this subsection. 

\section{Localization}\label{SL}

Localizing the incoming bosonic waves on the sky is crucial to understanding their microscopic production mechanism, which would open a new window in the multi-messenger astronomy with a quantum sensor network \cite{Dailey:2020sxa}. In this section, we follow the discussions in Sec.\,\ref{SFD} and evaluate the angular resolutions, which leads to the most optimal arrangement of the vector sensor pair.

Taking the momentum distribution in Eq.\,(\ref{str PDF}), we can calculate the asymptotic TS in Eq.\,(\ref{theta tilde}) and the corresponding Fisher information matrix in Eq.\,(\ref{var est}) in terms of $\hat{\Omega}_0$.
For simplicity, we consider a pair of detectors with the same noise spectrum $\lambda_1(\omega)=\lambda_2(\omega)= \lambda(\omega)$, and the same response function that stays constant $A_0$ within a narrow bandwidth $\Delta \omega$ around certain frequency $\omega_0$ and vanishes outside. In the case with only the longitudinal modes, the asymptotic TS  (\ref{theta tilde}) can be written as
\begin{equation}
\begin{aligned}
    &\frac{\tilde{\Theta} (\hat{\Omega}_0)}{\textrm{TS}_{\textrm{str}}^0} = 2\cos \left[ p_0 d\left( \hat{\Omega}_0^t-\hat{\Omega}_0\right)\cdot \hat{d} \right]M_{12}(\hat{\Omega}_0^t)\,M_{12}(\hat{\Omega}_0) \\
    &-M_{12}^2(\hat{\Omega}_0)+ \sum_{I=1,2}\left(M_{II}(\hat{\Omega}_0^t)-\frac{1}{2}M_{II}(\hat{\Omega}_0) \right)M_{II}(\hat{\Omega}_0), 
\end{aligned} \label{axionloc}
\end{equation}
for a truth value $\hat{\Omega}_0^t$,
where the normalization constant $\textrm{TS}_{\textrm{str}}^0$ is
\begin{equation}
    \textrm{TS}_{\textrm{str}}^0 \equiv \frac{ \rho^2\, T\, A_0^4 \,f_{\textrm{str}}^2(p_0) \,\zeta_\mathcal{O}^4(p_0)\, \omega_0^4\,\Delta \omega}{2 \, \pi \,\bar{\omega}^2 \, \lambda^2(\omega_0) \, p_0^2 \, m^4},
\end{equation} 
and the projection function $M_{IJ} (\hat{n})$ is defined as 
\begin{equation}
    M_{IJ}(\hat{n}) \equiv \left( \hat{n} \cdot \hat{l}_I \right) \left( \hat{n} \cdot \hat{l}_J \right).
\end{equation}
For the transverse modes distributed equally among right-hand and left-hand circular polarization, the asymptotic TS  (\ref{theta tilde}) is

\begin{equation}
\begin{aligned}
    &\frac{\tilde{\Theta} (\hat{\Omega}_0)}{\textrm{TS}_{\textrm{str}}^{R/L}}=2 \cos\left[ p_0 d\left( \hat{\Omega}_0^t-\hat{\Omega}_0\right)\cdot \hat{d} \right]\, \big( \hat{l}_1 \cdot \hat{l}_2-M_{12}(\hat{\Omega}_0^t)\big)\\
    &\times \big( \hat{l}_1 \cdot \hat{l}_2-M_{12}(\hat{\Omega}_0)\big)-\big( \hat{l}_1 \cdot \hat{l}_2-M_{12}(\hat{\Omega}_0)\big)^2\\
    &+\sum_{I=1,2}\left(  \frac{1}{2}-M_{II}(\hat{\Omega}_0^t)+\frac{1}{2} M_{II}(\hat{\Omega}_0)\right)\big( 1-M_{II}(\hat{\Omega}_0)\big)
\label{mdmloc}
\end{aligned}
\end{equation}
where we define $\textrm{TS}_{\textrm{str}}^{R/L}$ as
\begin{equation}
    \textrm{TS}_{\textrm{str}}^{R/L} \equiv \frac{  \rho^2\, T \,A_0^4\,f_{\textrm{str}}^2(p_0) \,\zeta_\mathcal{O}^4(p_0)\, \Delta \omega}{8 \, \pi \,\bar{\omega}^2 \, \lambda^2(\omega_0) \, p_0^2  }.\label{TSRL}
\end{equation}

Due to the complicated forms of Eq.\,(\ref{axionloc}) and (\ref{mdmloc}), we will divide the discussions into two cases in the following subsections, in terms of the ratio between the baseline length and the de Broglie wavelength, i.e., $p_0 d$. The short baseline corresponds to $p_0 d \ll 1$ while the long-baseline cases are when $p_0 d \gg 1$.

\subsection{Short Baseline}
In the short-baseline limit, the pair of vector sensors are effectively located at the same place. The localization comes purely from the relative amplitude of the signals between the two vector sensors with different sensitive directions, as mentioned in \cite{Dailey:2020sxa}.

We define the angular resolution of $\hat{p}_0$ as the solid-angle uncertainty \cite{Cutler:1997ta,Graham:2017lmg}
\begin{equation}
    \sigma^2_\Omega \equiv \frac{2\pi \sin \theta^t}{ \sqrt{I_{\theta \theta}I_{\phi\phi}-I_{\theta\phi}^2}},\label{SAdef}
\end{equation}   
where spherical angles $\theta$ and $\phi$  parametrize $\hat{\Omega}_0$ in a specific coordinate, and the components of the Fisher information matrix can be calculated through Eq.\,(\ref{var est}).

\begin{figure}[htb]
    \centering
    \includegraphics[width=0.8\columnwidth]{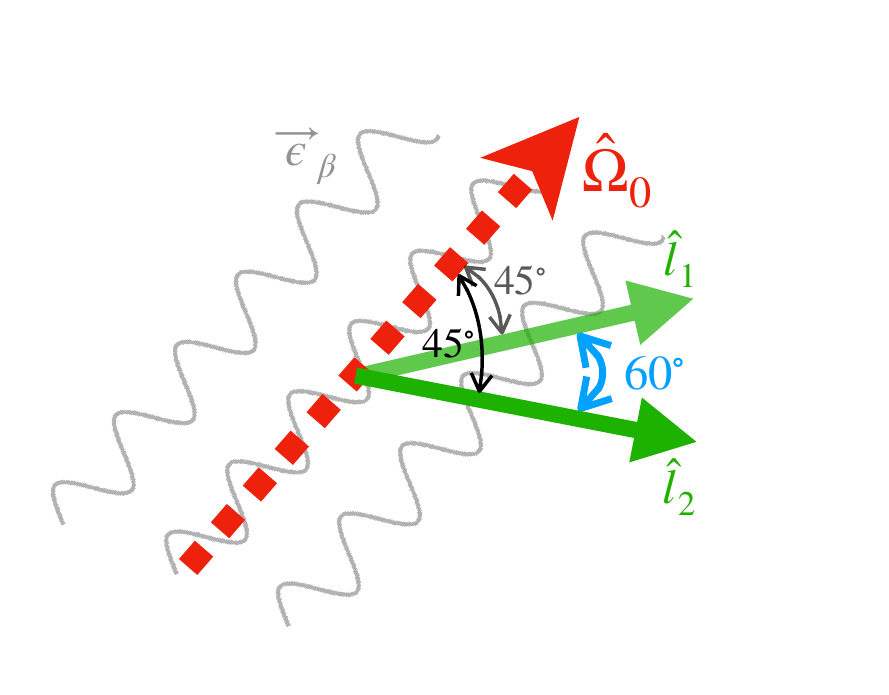}
    \caption{{ A schematic diagram is shown for the optimal sensitive directions of the two vector sensors in green arrows at the same location to localize the bosonic waves from a specific direction. The gray wavy lines represent the bosonic waves, and the red dashed arrow represents their incoming direction. }}
    \label{SBL}
\end{figure}

One chooses the coordinate system where the $\hat{z}$-axis is perpendicular to the plane that $\hat{l}_1$ and $\hat{l}_2$ span such that $\theta_{l_1} = \theta_{l_2}=\pi/2$. In this system, the solid-angle uncertainty (\ref{SAdef}) for the longitudinal-mode-only case is
\begin{equation}
\begin{aligned}
    \sigma^2_\Omega=&\frac{2\pi}{\textrm{TS}_{\textrm{str}}^0} \, \frac{1}{\sin^2 \theta^t \,|\cos \theta^t|}\\
    \times&\frac{1}{\left( \cos^2\left[\phi_{l_1}-\phi^t\right]+\cos^2\left[\phi_{l_2}-\phi^t\right] \right)\,|\sin[\phi_{l_1}-\phi_{l_2}]|} \\
    \geq & \frac{4\pi}{\textrm{TS}_{\textrm{str}}^0},\label{SAUL}
\end{aligned}
\end{equation}
where the minimization happens when $\phi_{l_1}-\phi^t=-(\phi_{l_2}-\phi^t)=\pi/6$ or $5\pi/6$ and $\cos \theta^t=\pm 1/\sqrt{3}$, as shown in Fig.\,\ref{SBL}. In such case the angle between $\hat{l}_1$ and $\hat{l}_2$ is $\pi/3$ and the one between $\hat{\Omega}_0^t$ and $\hat{l}_1$ or $\hat{l}_2$ is $\pi/4$.

For the transverse modes without macroscopic polarization, the solid-angle uncertainty has the same form as Eq.\,(\ref{SAUL}), except for replacing $\textrm{TS}_{\textrm{str}}^0$ with $\textrm{TS}_{\textrm{str}}^{R/L}$ defined in Eq.\,(\ref{TSRL}), as well as the same minimization condition, since the Fisher information matrix derived from Eq.\,(\ref{FFD}) and (\ref{FRFL}) is the same up to a normalization factor.

\subsection{Long Baseline}
In the long-baseline case, the Fisher information matrix is dominated by the term proportional to $p_0 d$. When the direction of the baseline $\hat{d}$ is along the $\hat{z}$-axis of the coordinate system, the information extracted from this single baseline is the polar angle $\theta$, which is the same as the scalar sensor interferometry \cite{Foster:2020fln}. The whole solid-angle can be localized by an array consisting of multiple baselines, similar to the Very Long Baseline Interferometry technique used by the Event Horizon Telescope \cite{EventHorizonTelescope:2019uob} in radio astronomy. Additionally, the self-rotation of the Earth can also change the direction of the baseline, serving the same purpose \cite{Foster:2020fln}. 

More explicitly, the order of $I_{\theta \theta}$ is much higher than that of $I_{\phi\phi}$ or $I_{\theta \phi}$ in terms of $p_0 d$. Thus we approximate the polar angle resolution as $\sigma_\theta^2 \approx I_{\theta \theta}^{-1}$, which leads to different results between the longitudinal modes and the transverse ones. For the longitudinal-mode-only case, $\sigma_\theta^2$ is
\begin{equation}
\begin{aligned}
   \sigma_\theta^2 \approx I_{\theta \theta}^{-1} &=\frac{1}{\textrm{TS}_{\textrm{str}}^0}\,\frac{1}{p_0^2 d^2\,\sin^2 \theta^t}\, \frac{1}{M_{12}(\hat{\Omega}_0^t)^2} \\
   &\geq \frac{1}{\textrm{TS}_{\textrm{str}}^0}\,\frac{1}{p_0^2 d^2\,\sin^2 \theta^t},\label{sigmatheta0}
\end{aligned}
\end{equation}
which is minimized when both $\hat{l}_1$ and $\hat{l}_2$ point towards $\hat{\Omega}_0^t$. We leave the term $1/\sin^2 \theta^t$ since the baseline direction $\hat{d}$ may be hard to tune, otherwise it can be further optimized at the transverse directions to $\hat{\Omega}_0^t$.

On the other hand, for the transverse modes without macroscopic polarization, $\sigma_\theta^2$ becomes
\begin{equation}
\begin{aligned}
    \sigma_\theta^2 \approx I_{\theta \theta}^{-1} =&\frac{1}{\textrm{TS}_{\textrm{str}}^{R/L}}\,\frac{1}{p_0^2 d^2\,\sin^2 \theta^t}\, \frac{1}{\left(\hat{l}_1 \cdot \hat{l}_2 -M_{12}(\hat{\Omega}_0^t) \right)^2} \\
    \geq& \frac{1}{\textrm{TS}_{\textrm{str}}^{R/L}}\,\frac{1}{p_0^2 d^2\,\sin^2 \theta^t},\label{sigmathetaRL}
\end{aligned}
\end{equation}
which leads to a similar form of the optimal resolution to that of the longitudinal modes in Eq.\,(\ref{sigmatheta0}), however with a different minimization condition, i.e., $\hat{l}_1 \,\parallel \,\hat{l}_2 \perp \hat{\Omega}_0^t$.

\begin{figure}[htb]
    \centering
    \includegraphics[width=0.75\columnwidth]{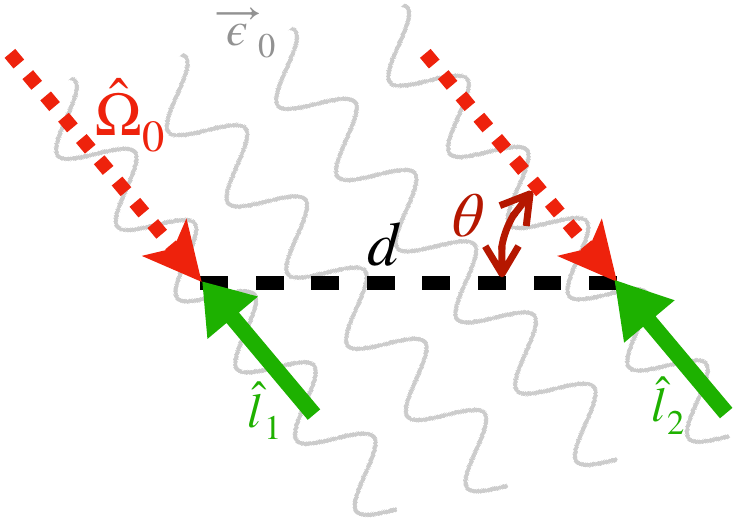}    \includegraphics[width=0.75\columnwidth]{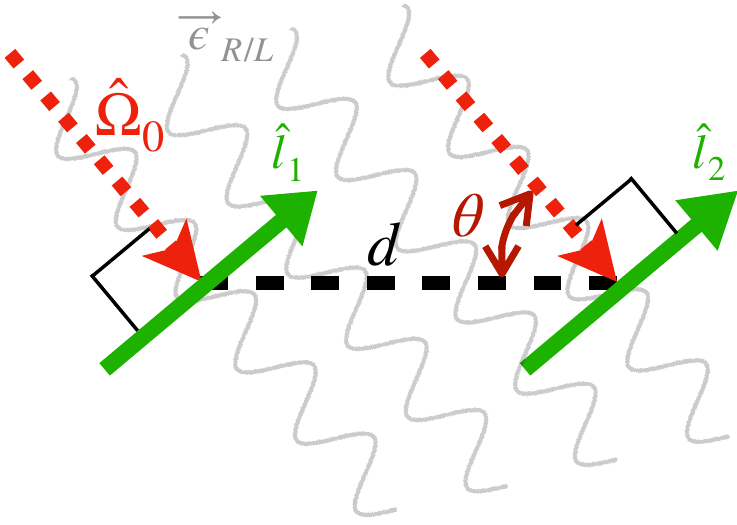}
    \caption{{
      Two schematic diagrams are shown for the optimal sensitive directions of the two vector sensors in green arrows separated by the baseline $d$ in black dashed lines to localize the bosonic waves from a specific direction. The gray wavy lines represent the bosonic waves, and the red dashed arrow represents their incoming direction. The upper panel is for the longitudinal modes only, while the lower panel is for the transverse modes without macroscopic polarization. The sensitive directions are set to make the projections of the signals on them as large as possible.
     }}
    \label{LBL}
\end{figure}

We show the optimal arrangements of the vector sensor pair with the long baseline in Fig.\,\ref{LBL}. Apparently, the optimized condition for the sensitive directions $\hat{l}$ is to make the projections of the signals on them as large as possible.

\begin{figure*}[htb]
    \centering
    \includegraphics[width=0.8\columnwidth]{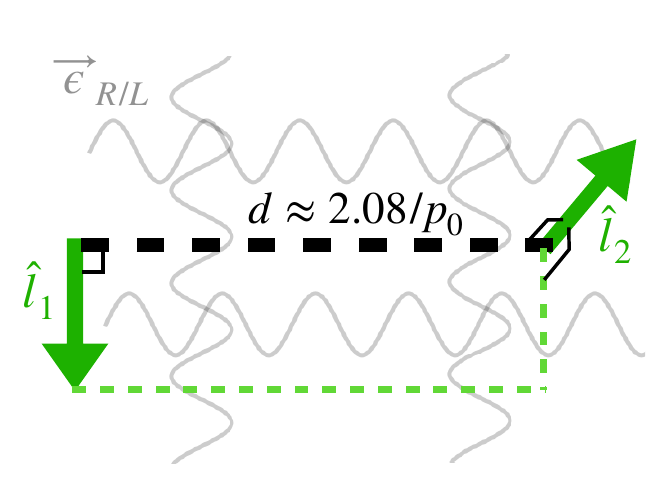}\qquad
    \includegraphics[width=0.8\columnwidth]{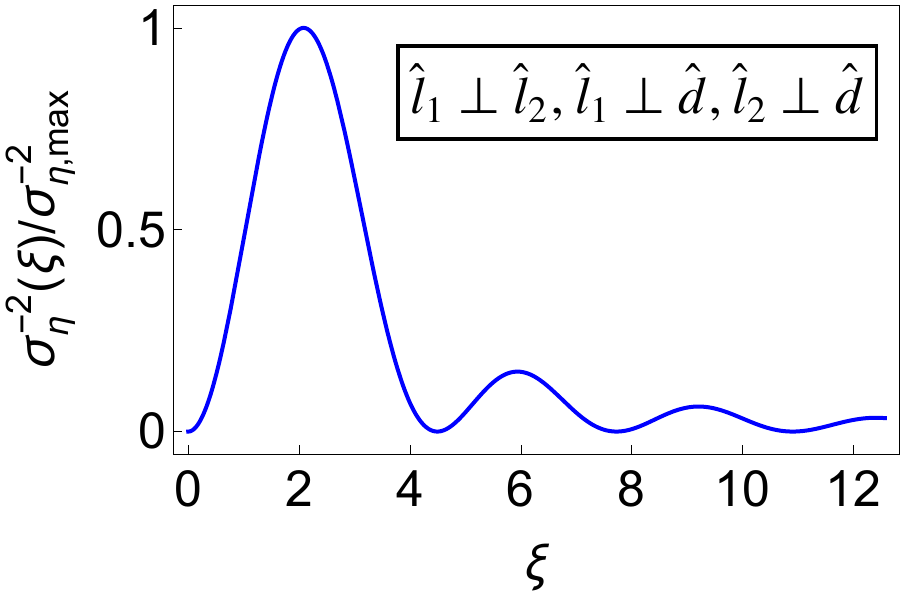}
    \caption{{
        A schematic diagram is shown in the left panel for the optimal sensitive directions of the two vector sensors separated by the baseline $d$ to detect the macroscopic circular polarization of isotropic bosonic waves, where the three directions are required to be perpendicular to each other. The green arrows show the sensitive directions, the black dashed line is the baseline, and the gray wavy lines represent the bosonic backgrounds.
    The right panel shows  $\sigma_\eta^{-2}$ as a function of $\xi \equiv p_0 d$ according to  Eq.\,(\ref{sigmaeta}). }}
    \label{fig:gpolar}
\end{figure*}

\section{Chiral Dark Photon Background}\label{CP}

For a cosmological dark photon background with an isotropic momentum distribution, we can expect a macroscopic circular polarization if the production mechanism is parity-violating, such as the tachyonic instability induced from the oscillation of a pseudo-scalar \cite{Anber:2009ua}. This section shows how to detect such a chiral signal with a pair of vector sensors separated by a finite long baseline.

We first introduce a new parameter 
\be \eta \equiv h(R)-h(L)\ee
to parametrize the difference between the right-hand and the left-hand circular polarization modes. Thus  Eq.\,(\ref{dipf2}) can be rewritten as
\begin{equation}
\begin{aligned}
    &F_{IJ}(\omega, \vec{d}, \hat{l}_I,\hat{l}_J)\\
    =\, & \frac{1}{2} \, \zeta^2_\mathcal{O}(p)\,\frac{f_{\textrm{iso}}(p)}{p}\,\Big( I_{IJ}(\omega, \vec{d}, \hat{l}_I,\hat{l}_J)+\eta V_{IJ}(\omega, \vec{d}, \hat{l}_I,\hat{l}_J) \Big),\label{FCP}
\end{aligned}
\end{equation}
where
\begin{equation}
\begin{aligned}
   &I_{IJ}(\omega, \vec{d}, \hat{l}_I,\hat{l}_J) \\
   \equiv \,& \frac{p^2}{f_{\textrm{iso}}(p)} \Big(\mathcal{F}^{R}_{IJ} (\omega, \vec{d}, \hat{l}_I,\hat{l}_J) +\mathcal{F}^{L}_{IJ} (\omega, \vec{d}, \hat{l}_I,\hat{l}_J) \Big),
\end{aligned}
\end{equation}
\begin{equation}
\begin{aligned}
   &V_{IJ}(\omega, \vec{d}, \hat{l}_I,\hat{l}_J) \\
   \equiv \,& \frac{p^2}{f_{\textrm{iso}}(p)} \Big(\mathcal{F}^{R}_{IJ} (\omega, \vec{d}, \hat{l}_I,\hat{l}_J) -\mathcal{F}^{L}_{IJ} (\omega, \vec{d}, \hat{l}_I,\hat{l}_J)  \Big) .
\end{aligned}
\end{equation}
%\begin{align}
%    I_{IJ}(\omega, \vec{d}, \hat{l}_I,\hat{l}_J) \equiv \frac{p^2}{f_{\textrm{iso}}(p)} \Big(\mathcal{F}^{R}_{IJ} (\omega, \vec{d}, \hat{l}_I,\hat{l}_J) +\mathcal{F}^{L}_{IJ} (\omega, \vec{d}, \hat{l}_I,\hat{l}_J) \Big),\\
%    V_{IJ}(\omega, \vec{d}, \hat{l}_I,\hat{l}_J) \equiv \frac{p^2}{f_{\textrm{iso}}(p)} \Big(\mathcal{F}^{R}_{IJ} (\omega, \vec{d}, \hat{l}_I,\hat{l}_J) -\mathcal{F}^{L}_{IJ} (\omega, \vec{d}, \hat{l}_I,\hat{l}_J)  \Big) .
%\end{align}
%and the labels $\hat{d}, \hat{l}_{I/J}$ are omitted for simplicity.
Notice that the dimensionless function $I_{IJ}$ and $V_{IJ}$ are defined in the analogy of the Stokes parameters $I$ and $V$ in radio astronomy.

According to Eq.\,(\ref{isoLR}), $V_{IJ}$ vanishes if the two vector sensors are at the same location. Therefore, to measure $\eta$ in Eq.\,(\ref{FCP}), one must have a finite long baseline $\vec{d}$. Using the same assumption as in Sec.\,\ref{SL} that the two detectors share the same noise spectrum $\lambda(\omega)$ and the constant response $A_0$ within $\Delta \omega$, the asymptotic TS in Eq.\,(\ref{theta tilde}) takes the form
\begin{equation}
\begin{aligned}
    \frac{\tilde{\Theta} (\eta)}{\textrm{TS}_{\textrm{iso}}^{R/L}} =&\, \frac{1}{2}\left( I_{11}^2+I_{22}^2+2I_{12}I_{21}\right)+\eta^t\left( V_{12}I_{21}+V_{21}I_{12}\right)\\
    +&\,V_{21}V_{12}\left(2\eta^t\eta-\eta^2\right),
    \label{PolarizationTS}
\end{aligned}
\end{equation}
where $\eta^t$ is the truth value of $\eta$, and
$\textrm{TS}^{R/L}_{\textrm{iso}}$ only differs from $ \textrm{TS}_{\textrm{str}}^{R/L}$ in Eq.\,(\ref{TSRL}) in the momentum distribution
\begin{equation}
    \textrm{TS}^{R/L}_{\textrm{iso}} \equiv \frac{  \rho^2\, T\,A_0^4\, f_{\textrm{iso}}^2(p_0)\, \zeta_\mathcal{O}^4(p_0)\, \Delta \omega}{8\, \pi \,\bar{\omega}^2\, \lambda^2(\omega_0)\,p_0^2}.
\end{equation}  

From Eq.\,(\ref{PolarizationTS}) and Eq.\,(\ref{var est}), the uncertainty of $\eta$ can be evaluated as
\begin{equation}
\begin{aligned}
    \sigma_{\eta}^2&=\frac{1}{\textrm{TS}^{R/L}_{\textrm{iso}} } \frac{1}{\left(\sin \theta_{l_1} \sin \theta_{l_2} \sin \left[\phi_{l_1}-\phi_{l_2} \right]\right)^2} \frac{\xi^4}{\left(\xi \cos \xi -\sin \xi\right)^2} \\
    &\geq \frac{5.26}{\textrm{TS}^{R/L}_{\textrm{iso}}},
    \label{sigmaeta}
\end{aligned}
\end{equation}
in the coordinate system where $\hat{d}$ is along the $\hat{z}$-axis. 
The minimization condition for Eq.\,(\ref{sigmaeta}) requires that $\theta_{l_1} = \theta_{l_2} = \pi/2$ and $\phi_{l_1} - \phi_{l_2} = \pi/2$, i.e., $\hat{l}_1, \hat{l}_2$ and $\hat{d}$ are perpendicular to each other. The arrangement of the sensor directions is shown in the left panel of Fig.\,\ref{fig:gpolar}. The dependence on $\xi \equiv p_0 d$ is shown in the right panel of Fig.\,\ref{fig:gpolar}, from which the most optimal separation is $\xi  \approx 2.08$.

\section{Conclusions} \label{conclusion}
In this work, we explore several applications of an array of vector sensors in identifying the features of bosonic waves, including angular distribution, types of coupling, directions of the emission, and macroscopic circular polarization.  Such information can be extracted from a pair of vector sensors with different optimal sensitive directions and a baseline that separates them for various purposes. These applications are based on two-point correlation functions of the vector fields with two spatial component labels.

We show that, for isotropic sources with a cosmological origin, one always gets the universal dipole angular correlation for all the polarization modes with two vector sensors at the same location. Any deviation from this relation is a signal of anisotropy. We find that a finite long baseline is necessary to break the degeneracy between the longitudinal and the transverse modes, leading to different correlation signals for the two polarization modes when the sensitive directions are parallel with the baseline or perpendicular to it. One can further identify a potential macroscopic circular polarization by making the two sensitive directions perpendicular to the baseline and each other.

Longitudinal and transverse polarization degrees of freedom contribute differently to the angular correlations for sources from a specific direction. However, we find that the optimal sensitive directions of the vector sensors, in terms of the angular resolution, are the same in both cases. With a short baseline, the sensitivity reaches an order of the inverse of the signal-to-noise ratio. In the long-baseline limit, one can resolve the polar angle with a resolution of the order $p_0 d$, requiring the sensitive directions to overlap with the target vector signals as much as possible.

Finally, we show the spatial and angular correlations of the virialized dark matter with different vector-like couplings to the standard model, such as the axion-fermion couplings and the dipole couplings between dark photons and fermions. The axion and the dark photon coupled to the magnetic dipole momentum (MDM) of the fermions show completely different spatial correlations since the axion gradient signals are purely longitudinal while only the transverse modes interact with the fermions' spins for the MDM dark photon. Due to the movement of the Earth in the galaxy frame, one expects to see an anisotropic part in their angular correlations, which have different signs between the axion and the MDM dark photon. The $U(1)_{B-L}/U(1)_B$ dark photon, the kinetic-mixing dark photon, and the electric dipole momentum dark photon, on the other hand, contribute equally among all the polarization modes, showing spatial correlations similar to the correlations between two scalar signals and no anisotropic term in the angular correlations.

In practice, once a vector sensor sees a suspicious signal, the rest of the detectors in the world can either falsify it or identify the sources' properties by correlating the signals of different detectors, in the way that we discuss in this study. Complete extraction of the data gives rich information on the microscopic nature and the origin of the sources.

A possible future extension of this work is to develop optimal data analysis strategies to increase the scan rate of the wave-like dark matter or other stochastic backgrounds, with vector-like couplings to the standard model, and to employ these in vector sensor networks, such as the GNOME.

\section*{Acknowledgements} 
We are grateful to Yonatan Kahn for the lectures and useful discussions in the second international school on Quantum Sensors for Fundamental Physics, Huayang Song for reading and comments for the manuscript, and Yue Zhao for useful discussions. 
This work is supported by the National Key Research and Development Program of China under Grant No. 2020YFC2201501. 
Y.C. is supported by the China Postdoctoral Science Foundation under Grant No. 2020T130661, No. 2020M680688, the International Postdoctoral Exchange Fellowship Program, and by the National Natural Science Foundation of China (NSFC) under Grants No. 12047557.
M.J. is supported by the National Natural Science Foundation of China under Grants No. 12004371.
J.S. is supported by the National Natural Science Foundation of China under Grants No. 12025507, No. 12150015, No.12047503; and is supported by the Strategic Priority Research Program and Key Research Program of Frontier Science of the Chinese Academy of Sciences under Grants No. XDB21010200, No. XDB23010000, and No. ZDBS-LY-7003 and CAS project for Young Scientists in Basic Research YSBR-006.
X.X. is supported by  Deutsche Forschungsgemeinschaft under Germany’s Excellence Strategy EXC2121 “Quantum Universe” - 390833306.
Y.C., X.X. and Y.Z. would like to thank USTC for their kind hospitality.

\appendix
\renewcommand{\theequation}{A\arabic{equation}}
  \setcounter{equation}{0}

\section*{Appendix: Scalar Correlation and Ensemble Average}
This Appendix briefly reviews the two-point correlations of the stochastic scalar fields developed in \cite{Foster:2017hbq,Derevianko:2016vpm,Foster:2020fln}.

The bosonic fields can couple to the standard model particles in different ways. For dark matter, the coherently oscillating waves lead to AC signals in the observable sectors with frequency nearly equal to the boson mass. For a light $\phi$ coupled with the standard model particles in a linear portal, the induced signal $S$ in the detector is proportional to the field value $\phi$. Thus the correlations between the two detectors' signals lead to two-point correlation functions of the scalar fields at different times and locations,
\begin{equation}
    \cor{S_1 (t_1,\vec{x}_1)  S_2 (t_2,\vec{x}_2)}\propto  \cor{\phi(t_1,\vec{x}_1)\phi(t_2,\vec{x}_2)}. \label{signalcorr}
\end{equation}
To calculate the two-point correlation functions of a stochastic scalar field, one can either define a density matrix which is diagonal in momentum space \cite{Derevianko:2016vpm}, or parametrize the field as a superposition of $N_b$ plane waves, \cite{Foster:2017hbq}
\begin{equation}
    \phi(t,\vec{x})=\sum_{j=1}^{N_b} \sqrt{\frac{2 \rho_{\phi}}{N_b \bar{\omega}\omega_j}} \cos(\omega_j t-\vec{p}_j \cdot \vec{x}+\alpha_j)
    \label{random phase}
\end{equation}
where the relative phase $\alpha_j$ is uniformly distributed within $[0,2\pi)$, and the momentum $\vec{p}_j$ comes with the probability distribution function $f(\vec{p})$. 
One next defines the ensemble average on any observable $\mathcal{A}$ as
\begin{equation}
    \cor{ \mathcal{A} }\equiv \prod_{m,n=1}^{N_b} \int d^3 \vec{p}_m\, f(\vec{p}_m)\, \int_{0}^{2\pi}\,\frac{d \alpha_n}{2\pi}\, \mathcal{A},
    \label{ensemble}
\end{equation}
with which one can calculate Eq.\,(\ref{signalcorr}) as
\begin{equation}
    \begin{aligned}
         &\cor{\phi(t_1,\vec{x}_1)\phi(t_2,\vec{x}_2)} \\
         = &\prod_{m,n=1}^{N_b}\, \iint d^3 \vec{p}_m\, d \alpha_n\, 
         \sum_{i,j=1}^{N_b}\, \frac{2 f(\vec{p}_m) \rho_{\phi}}{2\pi N_b\, \bar{\omega}\sqrt{\omega_i \omega_j}}\,\\& \cos\left[\omega_i t_1-\vec{p}_i \cdot \vec{x}_1+\alpha_i\right]\, 
         \cos\left[\omega_j t_2-\vec{p}_j \cdot \vec{x}_2+\alpha_j\right] \\
         =&\prod_{m=1}^{N_b}\, \int d^3 \vec{p}_m\, \sum_{i,j=1}^{N_b}\, \frac{ f(\vec{p}_m)\,\rho_{\phi}}{N_b\, \bar{\omega}\sqrt{\omega_i \omega_j}}\\&
         \cos\left[\left(\omega_i t_1-\omega_j t_2\right)-\left(\vec{p}_i \cdot \vec{x}_1-\vec{p}_j \cdot \vec{x}_2 \right) \right] \, \delta_{ij}\\
         =&\prod_{m=1}^{N_b} \int d^3 \vec{p}_m \,\sum_{i=1}^{N_b}\frac{ f(\vec{p}_m)\, \rho_{\phi}}{N_b\, \bar{\omega}\, \omega_i} \cos\left[\omega_i \tau-\vec{p}_i \cdot \vec{d}\, \right]\\
         =&\frac{ \rho_{\phi}}{\bar{\omega}}\int  \frac{d^3 \vec{p}\,f(\vec{p})}{\omega} \cos\left[ \omega \tau-\vec{p} \cdot \vec{d}\, \right],
    \end{aligned}
    \label{corr}
\end{equation}
where $\tau=t_1-t_2$, $\vec{d}=\vec{x}_1-\vec{x}_2$.

\begin{figure}[htb]
    \centering
    \includegraphics[width=0.9\columnwidth]{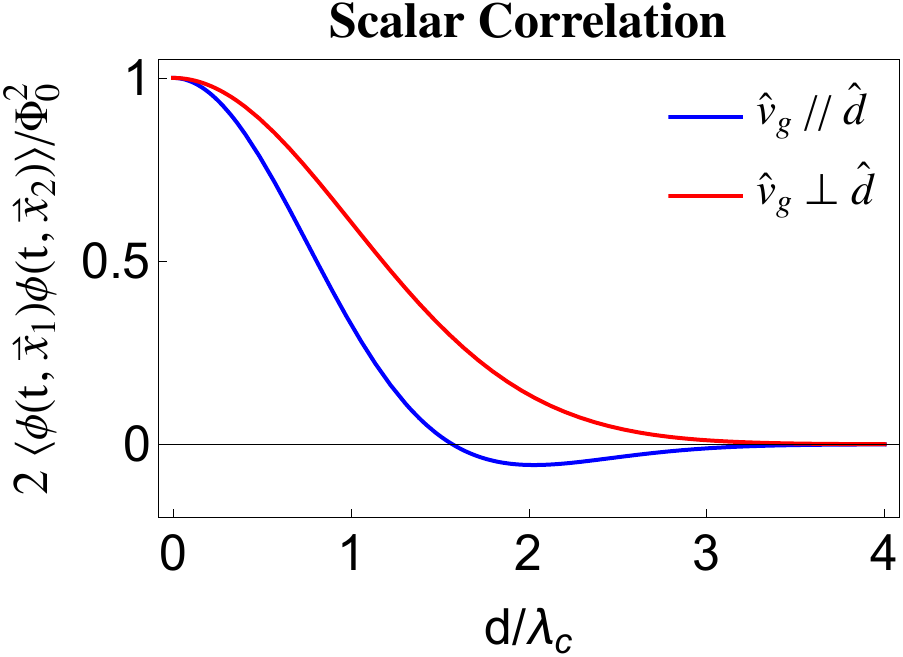}
    \caption{{The two-point correlation functions of scalar-like signals of the dark matter in Eq.\,(\ref{scalar}) is shown, which is identical to the one of the EDM dark photon in Fig.\,\ref{fig:edm} up to a normalization factor. }} 
    \label{fig:mono}
\end{figure}

Taking standard halo model with velocity distribution in Eq.\,(\ref{fDM}), one can evaluate the equal-time correlations of Eq.(\ref{corr}) as
\begin{equation}
\begin{aligned}
   \cor{\phi(t,\vec{x}_1)\phi(t,\vec{x}_2)} =\frac{\Phi_0^2}{2} \exp \left(-\frac{d^2}{2 \lambda_c^2}\right)\cos\left( m\, \vec{v}_g\cdot \vec{d} \right),
\end{aligned}
    \label{scalar}
\end{equation}
where $\Phi_0 \equiv \sqrt{2 \rho_{\phi}}/\bar{\omega}$, $\lambda_c \equiv 1/\left(m\, \vvir\right)$, and $\vvir$ and $v_g$ are the virial velocity and the Earth's velocity in the galactic reference frame respectively.  The spatial correlation (\ref{scalar}) is shown in Fig.\,\ref{fig:mono}, which is the same as that for the EDM dark photon in Fig.\,\ref{fig:edm} up to a normalization factor.

For vector-like signals involving polarization modes $\beta$, one can generalize the ensemble average on any observable $\xi$ to be
\begin{equation}
    \cor{\xi} \equiv \prod_{i,j,k}^{N_b}\int d^3\vec{p}_i\, f(\vec{p}_i) \int_0^{2\pi}\,  \sum_{\beta_k}h(\beta_k)\, \frac{d \alpha_j}{2 \pi} \, \xi,
\end{equation}
where $h(\beta_k)$ is the percentage of the polarization mode $\beta_k$.

\newpage
\bibliography{references}

%\bibliographystyle{unsrt}
%\bibliography{references}

\end{document}